\documentclass[acmsmall, nonacm]{acmart} 

\AtBeginDocument{%
  \providecommand\BibTeX{{%
    \normalfont B\kern-0.5em{\scshape i\kern-0.25em b}\kern-0.8em\TeX}}}

\setcopyright{acmcopyright}
\copyrightyear{2021}
\acmYear{2021}
\acmDOI{10.1145/1122445.1122456}

\acmConference[Woodstock '18]{Woodstock '18: ACM Symposium on Neural
  Gaze Detection}{June 03--05, 2018}{Woodstock, NY}
\acmBooktitle{Woodstock '18: ACM Symposium on Neural Gaze Detection,
  June 03--05, 2018, Woodstock, NY}
\acmPrice{15.00}
\acmISBN{978-1-4503-XXXX-X/18/06}

\usepackage{tikz}

\begin{document}

\title[Separating Polarization from Noise]{Separating Polarization from Noise: Comparison and Normalization of Structural Polarization Measures}

\author{Ali Salloum}
\email{ali.salloum@aalto.fi}
\orcid{0000-0002-2381-6876}
\affiliation{%
  \institution{Aalto University}
  \streetaddress{Konemiehentie 2, 02150}
  \city{Espoo}
  \country{Finland}}

\author{Ted Hsuan Yun Chen} 
\email{ted.hsuanyun.chen@gmail.com}
\orcid{0000-0002-3279-8710}
\affiliation{%
  \institution{Aalto University and University of Helsinki}
  \streetaddress{Unioninkatu 37, 00170}
  \city{Helsinki}
  \country{Finland}
}

\author{Mikko Kivel\"a}
\email{mikko.kivela@aalto.fi}
\orcid{1234-5678-9012}
\affiliation{%
  \institution{Aalto University}
  \streetaddress{Konemiehentie 2, 02150}
  \city{Espoo}
  \country{Finland}
}

\renewcommand{\shortauthors}{Salloum, Chen, and Kivel\"a}

\begin{abstract}

Quantifying the amount of polarization is crucial for understanding and studying political polarization in political and social systems.
Several methods are used commonly to measure polarization in social networks by purely inspecting their structure. We analyse eight of such methods and show that all of them yield high polarization scores even for random networks with similar density and degree distributions to typical real-world networks. 
Further, some of the methods are sensitive to degree distributions and relative sizes of the polarized groups. We propose normalization to the existing scores and a minimal set of tests that a score should pass in order for it to be suitable for separating polarized networks from random noise. 
The performance of the scores increased by 38\%-220\% after normalization in a classification task of 203 networks.
Further, we find that the choice of method is not as important as normalization, after which most of the methods have better performance than the best-performing method before normalization.
This work opens up the possibility to critically assess and compare the features and performance of different methods for measuring structural polarization.


\end{abstract}

\ccsdesc[500]{Human-centered computing~Collaborative and social computing~Collaborative and social computing design and evaluation methods~Social network analysis}
\ccsdesc[500]{Human-centered computing~Collaborative and social computing design and evaluation methods~Social network analysis}
\ccsdesc[500]{Human-centered computing~Social network analysis}
\ccsdesc[300]{Applied computing~Sociology}

\keywords{network science, polarization, political polarization, computational social science, normalization, twitter, networks, sociology, community detection, clustering, statistical significance}

\maketitle

\section{Introduction}


Political polarization has long been a question in the social sciences \cite{baldassarri2008partisans,fiorina2008political}, with mainstream interest growing following observed political divisions in the 2000 and 2004 US Presidential elections \cite{fiorina2008political}. Polarization, which is generally understood in the social science literature as the division of individuals into coherent and strongly opposed groups based on their opinion of one or more issues \cite{fiorina2008political,dimaggio1996have}, has deleterious consequences for social systems. These undesirable outcomes include increased divisiveness and animosity \cite{mason2015disrespectfully}, policy gridlock \cite{jones2001party}, and even decreased political representation \cite{baldassarri2008partisans}. Recent sociopolitical challenges have further strengthened academic interest in this area, as political polarization has been associated with difficulties in resolving pressing societal issues such as climate change \cite{zhou2016boomerangs}, immigration and race relations \cite{hout2020immigration}, and recently the COVID-19 pandemic \cite{makridis2020real}. In the context of social computing and computer mediated communication, political polarization on social media has been shown to constrain communication and eases the spread of misinformation \citep{iandoli2021impact}. 

Extensive research efforts have been put toward mitigating political polarization in computer mediated communication. This body of work ranges from studies exploring the role of polarization in social media \citep{rabab2016measuring,ozer2019measuring,darwish2019quantifying,demszky2019analyzing,bright2018explaining,conover2011political,cossard2020falling} to platform design aimed at attenuating polarization \citep{bail2018exposure,nelimarkka2018social,nelimarkka2019re}. A fundamental requirement that underlies these efforts is that we are able to identify polarized topics and systems, and monitor changes over time. \textit{How will we know where intervention is required? And how will we know if our interventions are successful?} Much research has been done in the area of measuring political polarization. Traditional methods primarily relied on survey-based approaches, which tend to measure distributional properties of responses to questions on public opinion surveys, such as bimodality and dispersion \cite{dimaggio1996have}. With the increasing availability and richness of publicly-observable social data, recent work from the computational social science and network science fields have shifted polarization measurement in two new directions \citep{garimella2018quantifying}. 
The first area of work are content-based approaches \citep[e.g.][]{belcastro2020learning,demszky2019analyzing}, which have become widely used following developments in natural language processing tools that allow researchers to detect antagonistic positions between groups on social media.

Another fruitful area of work focuses on structural aspects of polarization inferred from network representations of social or political systems. These structural polarization measures tend to be based on the logic of identifying what would be observable features of systems that are characterized by polarization. Because polarization is a system-level phenomenon, these features are defined at the network-level, making them different from node-level (i.e. individual) behavioral mechanisms. While node-level mechanisms such as homophily can contribute to polarization \citep{baumann2020modeling,jasny2015empirical}, they do not necessitate it, as individuals will at any time exhibit a multitude of behavioral tendencies. Most importantly, following the definition of polarization outlined above, structural measures generally take separation between structurally-identified communities to represent polarization between groups in the system. Additionally, different measures tend to be motivated by additional aspects of political polarization, such as, for example, the difficulty of information to escape echo chambers \citep[e.g.][]{garimella2018quantifying}. Because these structural polarization measures can flexibly capture theoretically-grounded aspects of political polarization, especially at a relatively low cost compared to content-based and survey-based approaches, they appear to be attractive measures for applied work.

Indeed, applications of structural polarization scores to real networks have spanned many domains, including  the studies of political party dynamics \cite{baldassarri2007dynamics, moody2013portrait, kirkland2014measurement, waugh2009party, neal2020sign}, cohesion and voting behavior \cite{waugh2009party, amelio2012analyzing, dal2014voting, chen2020polarization}, political events \cite{weber2013secular, morales2015measuring, darwish2019quantifying, cossard2020falling, rashed2020embeddings}, and online echo chambers \cite{garimella2018political, cota2019quantifying, cossard2020falling}. They have also been used to detect and study the presence of `controversy' in communication across different topics \cite{garimella2018quantifying}---in fact, some structural polarization scores are named as controversy scores---as polarized groupings can be interpreted as a manifestation of a controversial issue. Despite their widespread application, there are few systematic assessments of how well these structural polarization measures perform in key aspects of measurement validity \citep{trochim2001research}, such as predicting whether a network is polarized based on human labeling, and whether they are invariant to basic network statistics such as size and densiy. Exceptions to this include a small number of studies that compare the performance of some scores in classifying polarized political systems \cite{garimella2018quantifying,emamgholizadeh2020framework} or those that show certain scores are invariant to network size \cite{ozer2019measuring}. On the whole, the body of evidence is sparse. 

Further, beyond simply the lack of evidence, there are in fact theoretical reasons to expect systematic shortcomings with the current approach. Consider that the typical method for measuring structural polarization relies on the joint tasks of (1) graph clustering to find two groups, and then (2) measuring how separate the two groups are. A key characteristic of this approach is that most clustering methods start from the assumption that the networks have separate groups, and because these clustering algorithms are optimized to do so, they can find convincing group structures even in completely random networks \cite{zhang2014scalable, lancichinetti2010statistical, guimera2004modularity}. 
Moreover, the quality function of these methods can be sensitive to elementary features such as the size and density of these networks \cite{guimera2004modularity,zhang2014scalable}.
Given such a starting point, it is difficult to develop a measure that would separate non-polarized networks from polarized ones. Scores based on ostensibly intuitive community structures are potentially poor measures of polarization because they yield high polarization scores for non-polarized networks. 

We address these concerns by analyzing eight different scores that have been proposed and used for quantifying structural polarization. We test how well they separate random networks from polarized networks and how sensitive they are to various network features: number of nodes, average degree, degree distribution, and relative size of the two groups. Further, we compute these polarization values for a large array of political networks with varying degrees of polarization, and use null models to see how much of the score values are explained by these different network features. We find that none of the measures tested here are able to systematically produce zero scores (i.e., low or non-polarization) for random networks without any planted group structure, which we take to be a reasonable requirement for a polarization measure \cite{guerra2013measure}. Further, they are sensitive to variations in basic network features, with the average degree being particularly challenging for these measures. 

Our study makes a number of contributions to the literature. First, our results indicate that it is difficult to interpret any of the polarization scores in isolation. Given a score value it can be impossible to tell even whether the network is more polarized than a random network unless the score is reflected against the network features. Further, analysing the extent to which the network is polarized or making comparisons claiming that one network is more polarized than another one is not straightforward with the current measures. Second, we present a straightforward solution to the problem. We show that normalizing polarization scores against a distribution of scores from their corresponding configuration models improves their performance in a classification task of 203 labeled networks. Finally, we make our testing code and data publicly available on GitHub \cite{code1} and Zenodo \cite{code2}.

The rest of this paper is organized as follows. We first briefly discuss prior research and describe the current methods for measuring polarization in networks. Next, we introduce the methods and data used in this study. We then study the performance of modern polarization scores both on real networks and synthetic networks. Finally, we present normalization to the existing scores. We conclude by summarizing our results in the discussion section.


\section{Network-based Polarization Scores}

\begin{figure}[htb]
    \centering
    \tikzstyle{startstop} = [rectangle, rounded corners, minimum width=4cm, minimum height=0.8cm,text centered, draw=black, fill=gray!10]
    \tikzstyle{arrow} = [thick,->,>=stealth]
    \begin{tikzpicture}[node distance=4.75cm, font = \sffamily]
    \node (make) [startstop] {Construct Graph};
    \node (part) [startstop, right of=make] {Partition Graph};
    \node (calc) [startstop, right of=part] {Compute Polarization};
    \draw [arrow] (make) -- (part);
    \draw [arrow] (part) -- (calc);
    \end{tikzpicture}
    \caption{The polarization measurement pipeline}
    \label{fig:pipeline}
\end{figure}
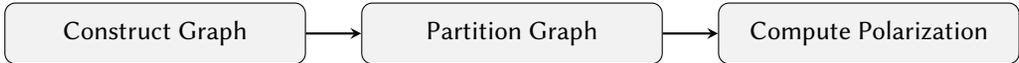

Polarization can be understood as the division of individuals into coherent and strongly opposed groups based on their opinion of one or more issues \cite{dimaggio1996have,fiorina2008political,mccoy2018polarization}. Polarization reflects strongly on social network structures by creating two internally well-connected groups that are sparsely connected to each other \cite{baumann2020modeling}, which means that the amount of polarization in a system can be measured by observing the structure of interactions within that system. This logic informs many structural polarization scores, which are obtained following the procedure shown in Fig. \ref{fig:pipeline}: (1) social network construction, (2) finding the two groups via graph partitioning, and (3) quantifying the separation between the two groups.

When measuring structural polarization in a social system, constructing a network representation of the system that allows us to identify the existence of polarized group structures apart from other types of interactions is a crucial step. It defines the appropriate bounds of the system, both in terms of which units are included and which interactions are measured. A social network is suitable for polarization analysis with the methods described here if the link between two nodes indicates a positive relationship, such as friendship, preference similarity, endorsement, or agreement \cite{conover2011political, garimella2018quantifying, akoglu2014quantifying, garimella2018thesis}. Bounding the network to the appropriate set of interactions is key, as it has been observed that simultaneously measuring unrelated topics will yield lower structural polarization scores regardless of how polarized each of the topics are \citep{chen2020polarization}.

Network-based polarization scores share the requirement that the network needs to be partitioned, with the logic being that the partitioned units are somehow closer to others in their in-group than to those from out-groups \cite{garimella2018quantifying}.
In some cases, information external to the network structure can be used to infer the groups in the network (e.g. node partisanship labels \cite{bright2018explaining}). However, the group membership or position of each unit is often not known or is difficult to otherwise infer, making graph partition algorithms necessary. Graph partitioning is an ill-posed problem \cite{fortunato2016community}, with a large number of algorithms available that try to find slightly different types of clusters in networks.
In this paper, we use three different partition algorithms: METIS \cite{karypis1998fast}, regularized spectral clustering \cite{zhang2018understanding}, and modularity \cite{newman2006modularity} optimization algorithm. 
These methods give similar overall results in our analysis (see Appendix \ref{C}); for brevity we will show only results for METIS, which searches for two large groups such that there is a minimum number of connections between them.

The main difference between structural polarization scores is how they compute the amount of polarization given a graph and its partition. As noted above, structural polarization between groups is measured by the separation between structurally-identified communities in the network. Here, separation is usually characterized by the discrepancy between interaction patterns that occur within groups and those that cross groups, but scores differ in the kind of interactions they highlight. In this study, we examine eight different structural polarization scores. A brief summary of all polarization scores examined in this study can be found in Table \ref{crouch}. We selected these scores because there is considerable variation in the kinds of interactions they highlight, and because they are measures that have been used in applied work across various fields, including computational social science \citep{darwish2019quantifying}, political science \citep{hargittai2008cross}, communications \citep{bright2018explaining}, and policy studies \citep{chen2020polarization}.

At its simplest, structural polarization is measured by comparing the difference in the frequency of external to internal links using the \textit{EI-index} \cite{krackhardt1988informal} and similar density-based scores \cite{chen2020polarization}. These scores disregard the internal structure of the groups and how the links between them are placed. The \textit{Betweeness Centrality Controversy} (BCC) score \cite{garimella2018quantifying} alternatively considers the difference in the edge betweenness centrality of external and internal links. Another approach is to use random walk simulations to determine how likely information is to remain within groups or reach external groups (i.e. \textit{Random Walk Controversy} and \textit{Adaptive Random Walk Controversy}; RWC and ARWC) \cite{garimella2018quantifying, rabab2016measuring, rumshisky2017combining, darwish2019quantifying}. We also explore the performance of a polarization measure based on community boundaries (\textit{Boundary Polarization}; BP) \cite{guerra2013measure}, where a high concentration of high-degree nodes in the boundary of communities implies lower polarization as the influential users are seen as bridging the gap between communities. Lastly, we study a measure based on label propagation (\textit{Dipole Polarization}; DP) \cite{morales2015measuring}, where the influence of high-degree nodes is believed to spread via the neighbors in the network, and the distance of the quantified influence of each community is proportional to the polarization. Full definitions of each of the eight structural polarization scores we study are included in Appendix \ref{A}. 

\begin{table}
    \caption{Polarization scores used in this study}
    \label{crouch}
    \begin{tabular}{ l l l }
        \toprule
\textbf{Polarization score}      
& \textbf{Domain}   
& \textbf{Parameters} \\\midrule
Random Walk Controversy (RWC) \cite{garimella2018quantifying}
& [-1, 1]       
& \# of influencers in each group \\
Adaptive Random Walk Controversy (ARWC)       
& [-1, 1]                        
& \% of influencers in each group  \\
Betweenness Centrality Controversy (BCC) \cite{garimella2018quantifying}        
& [0, 1] 
& Kernel for density estimation \\
Boundary Polarization (BP) \cite{guerra2013measure} &
[-0.5, 0.5]      
& -  \\
Dipole polarization (DP) \cite{morales2015measuring}       
& [0, 1]                        
& \% of influencers in each group  \\
E-I Index (EI) \cite{krackhardt1988informal}       
& [-1, 1] 
& - \\
Adaptive E-I Index (AEI) \cite{chen2020polarization} &
[-1, 1] 
& -  \\
Modularity (Q) \cite{waugh2009party}       
& [-0.5, 1] 
& - 
\\
        \bottomrule
    \end{tabular}
\end{table}

There are several possible issues with the approach to measuring structural polarization described above. First, if not designed carefully, such polarization scores can be very sensitive to ostensibly unrelated network features such as number of nodes, density, and degree heterogeneity. The scores can also behave in untransparent ways if the groups sizes are not balanced. Finally, graph partitioning algorithms can find clustering structures even in randomized networks, giving them very high polarization scores, especially if they are sparse. These are all problems that have already been found in related problem of partitioning networks to arbitrary number of clusters (as opposed to two) and evaluating the partition \cite{bagrow2012communities, zhang2014scalable, lancichinetti2010statistical, guimera2004modularity}. This kind of sensitivity to basic network features and cluster sizes 
means that the scores are not in an absolute scale where they could be compared across different networks
unless these confounding factors are taken into consideration.
Problems with polarization scores are not only a theoretical possiblity, but practical problems with the structural polarization score framework have been reported recently. The RWC score, which has been recently described as state-of-the-art \cite{emamgholizadeh2020framework, de2020measuring} and used as the main measure \cite{darwish2019quantifying, cossard2020falling, rashed2020embeddings}, has poor accuracy in separating polarized networks from non-polarized ones \cite{emamgholizadeh2020framework}.

\begin{figure}[htb]
    \centering
    \includegraphics[width=1\textwidth]{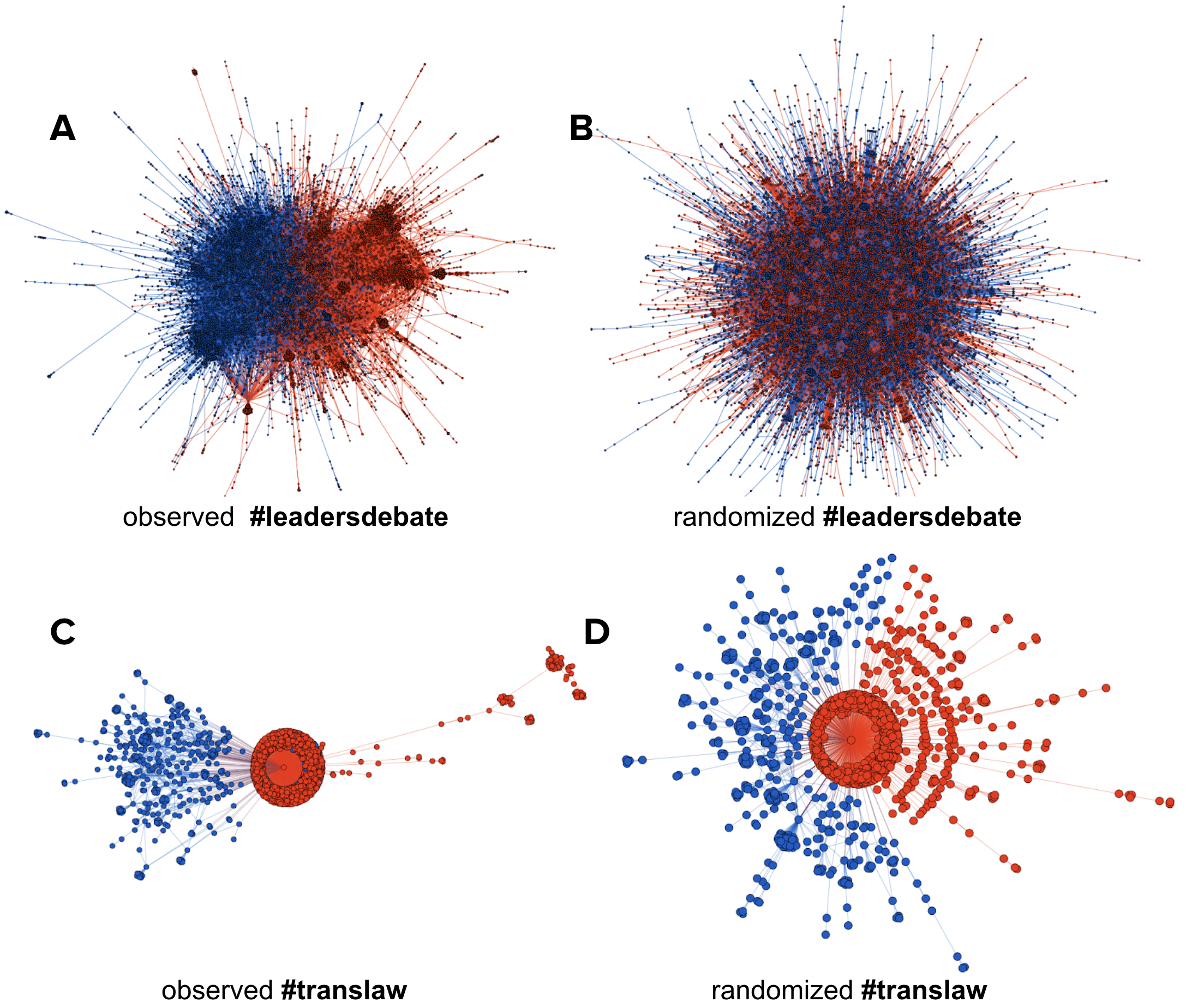}
    \caption{{\bf A} The network represents a Twitter conversation network for \#leadersdebate. Garimella et al. \cite{garimella2018quantifying} has labeled this topic polarized as it refers to the debate during the U.K. national elections in May 2015. {\bf C} A Twitter endorsement network around \#translaw during the Finnish Parliamentary Elections in 2019. ({\bf B}, {\bf D}) For both networks we also visualized their randomized versions obtained with configuration model. The following RWC values ($k=10$) were computed: observed \#leadersdebate 0.57, randomized \#leadersdebate 0.27,  observed \#translaw 0.68 and randomized \#translaw 0.74. RWC value of zero indicates no polarization and one indicates maximum polarization. Randomized networks still have positive polarization, and for the randomized \#translaw network the value is even higher than the observed network's score. Only the giant component of the network is visualized.}
    \label{fig:networkmat}
\end{figure}

We illustrate some of the problems in measuring structural polarization in practice using two networks (shown in Fig. \ref{fig:networkmat}).
The first network (\textit{\#leadersdebate}) demonstrates the general problem of all scores studied here. Here, the observed network has an RWC value of 0.57. After fixing the observed degree sequence of the network and randomizing everything else, the shuffled networks still scores positively, with an averaged RWC value of 0.27. 
That is, approximately half of the polarization score value is explained by the size and the degree distribution of the network. 

The second network (\textit{\#translaw}), with RWC value of 0.68, displays a serious problem related to hubs. A randomized network can have a higher polarization than the observed one due to the absorbing power of the hubs. In other words, a random network with one or more hubs can keep the random walker trapped inside the starting partition even in a non-polarized network. This leads to higher values of RWC as in the case of the \textit{\#translaw} network, where the randomized versions obtained an average score of 0.74. Note that this is also higher than the score for the first network, which has a clearly visible division to two clusters in the visualization. As will become evident in the following sections, this is likely due to the larger size and higher density of the first network. We will next systematically analyse how the various network features affect all of the eight structural polarization scores.

\section{Methods}
Our primary aim in this paper is to assess how well commonly-used structural polarization measures perform on important aspects of measurement validity \citep{trochim2001research}. We begin by examining the extent to which these eight structural polarization scores are driven by basic network properties using null models based on random networks.
These models are built in a way that they fix some structural properities but are otherwise maximally random, i.e., they give equal probability of sampling every network while holding that property constant. There are two main use cases. First, we will systematically analyse various network features by sweeping through the parameter spaces of these models. Second, we can match some of the features of the observed networks and randomize everything else, giving us an estimate of how much of the scores is a consequences of these structural features. Valid measures should not be systematically biased by these structural features.


We use (1) an Erd\H{o}s-R\'enyi model  \cite{erdos1959random} (fixing the number of links), (2) a configuration model \cite{molloy1995critical,fosdick2018configuring} (fixing the degree sequence), and (3) a model for fixing degree-degree sequences \cite{mahadevan2006systematic}. All of these models fix the number of nodes. One can see these models as a sequence where each of them shuffle less of the data than the previous one \cite{mahadevan2006systematic,gauvin2018randomized}, because fixing degree-degree sequences automatically fixes the degree distribution, which in turn fixes the number of links. 
To emphasize this relationship between the models, we label them using the $dk$-series framework \cite{mahadevan2006systematic, orsini2015quantifying}.
The $dk$-distribution of graph $G$ incorporates all degree correlations within $d$-sized subgraphs, hence allowing us to `shuffle' the original graph while keeping the desired prescribed properties regarding network topology. Increasing values of $d$ captures more properties of the network at the expense of more complex probability distributions. In the limit of large $d$, the complete representation of the input graph is obtained. The intuition is very similar to Taylor series. The more terms you include in the series (corresponding to higher values of $d$), the closer you get to the exact value of the function (i.e. the original network). 
Within this framework, the  Erd\H{o}s-R\'enyi network fixing the number of links (or equivalently, the average node degree) is $d=0$, the configuration model fixing the degree sequence is $d=1$, and the model fixing the degree-degree sequence is $d=2$. Each randomized network is repartitioned before computing its polarization score by applying the same graph partitioning algorithm as for the corresponding original network.


In addition to models fixing some observable network properties, we use the stochastic block model \cite{holland1983stochastic} to generate networks with two groups. This model is used in Section \ref{sbmsection} to explore how unbalanced group sizes affect the score values.

\subsection{Data}\label{datasection}

\begin{figure}[ht]
    \centering
    \includegraphics[width=1\textwidth]{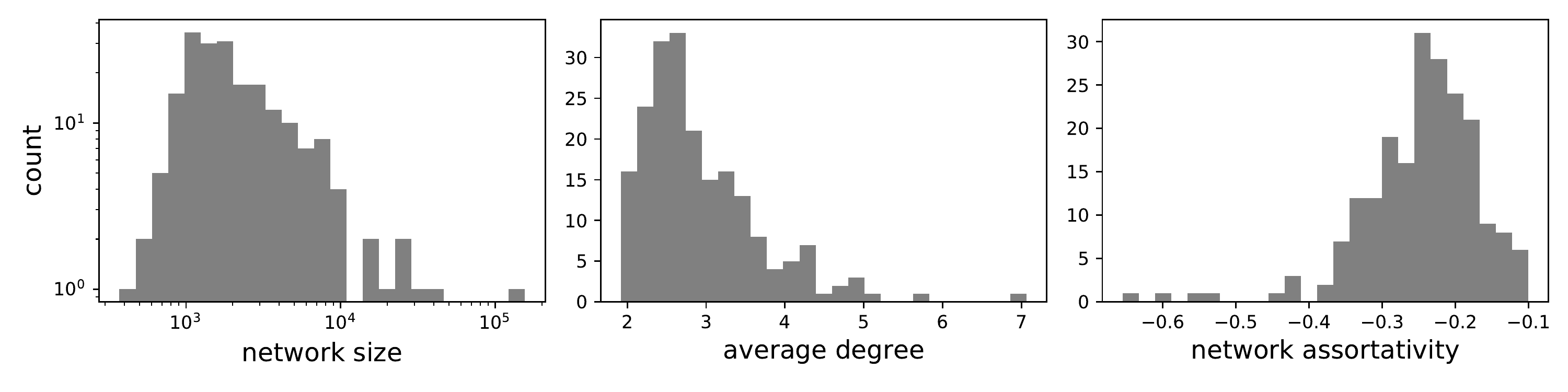}
    \caption{Distributons of basic statistics of the 203 observed networks. The average network studied had approximately 4000 nodes and an average degree of 3. Network assortativity refers to degree assortativity. 
    }
    \label{fig:summary}
\end{figure}

In addition to networks simulated with models, we use real-world data from Twitter from three sources for a total of 203 networks. First, from Chen et al., we used 150 networks constructed from the most popular hashtags during the 2019 Finnish Parliamentary Elections \cite{chen2020polarization}. These were constructed based on single hashtags (e.g. \textit{\#police, \#nature, \#immigration}). Second, from the same source, we included 33 topic networks, which were constructed from sets of hashtags related to larger topics, such as climate change or education \cite[see Appendix A in][]{chen2020polarization}. Third, we used 20 topic networks from Garimella et al.'s study on the RWC \citep{garimella2018quantifying}. Each of the 203 resulting network has a node set containing all users who posted an original tweet with a specific set of hashtags and all users who retweeted at least one of these tweets. Undirected ties on the network indicate the connected nodes have at least one instance of retweeting between them on the given topic. 

Finally, we process all network prior to assessment by (1) obtaining the giant component of the network as has been done previously \cite{garimella2018quantifying}, (2) removing self-loops, and (3) removing parallel edges. The latter two steps did not have a significant effect on polarization values. The average network in this study had approximately 4000 nodes, an average degree of 3 and tended to be slightly assortative. Complete summary distributions for the networks included in our study are presented in Fig. \ref{fig:summary}.



\section{Results}

\subsection{Real Networks}\label{rnetana}

\begin{figure}[htbp]
    \centering
    \includegraphics[width=1\textwidth, trim = {0.275cm 0 0.255cm 0}, clip]{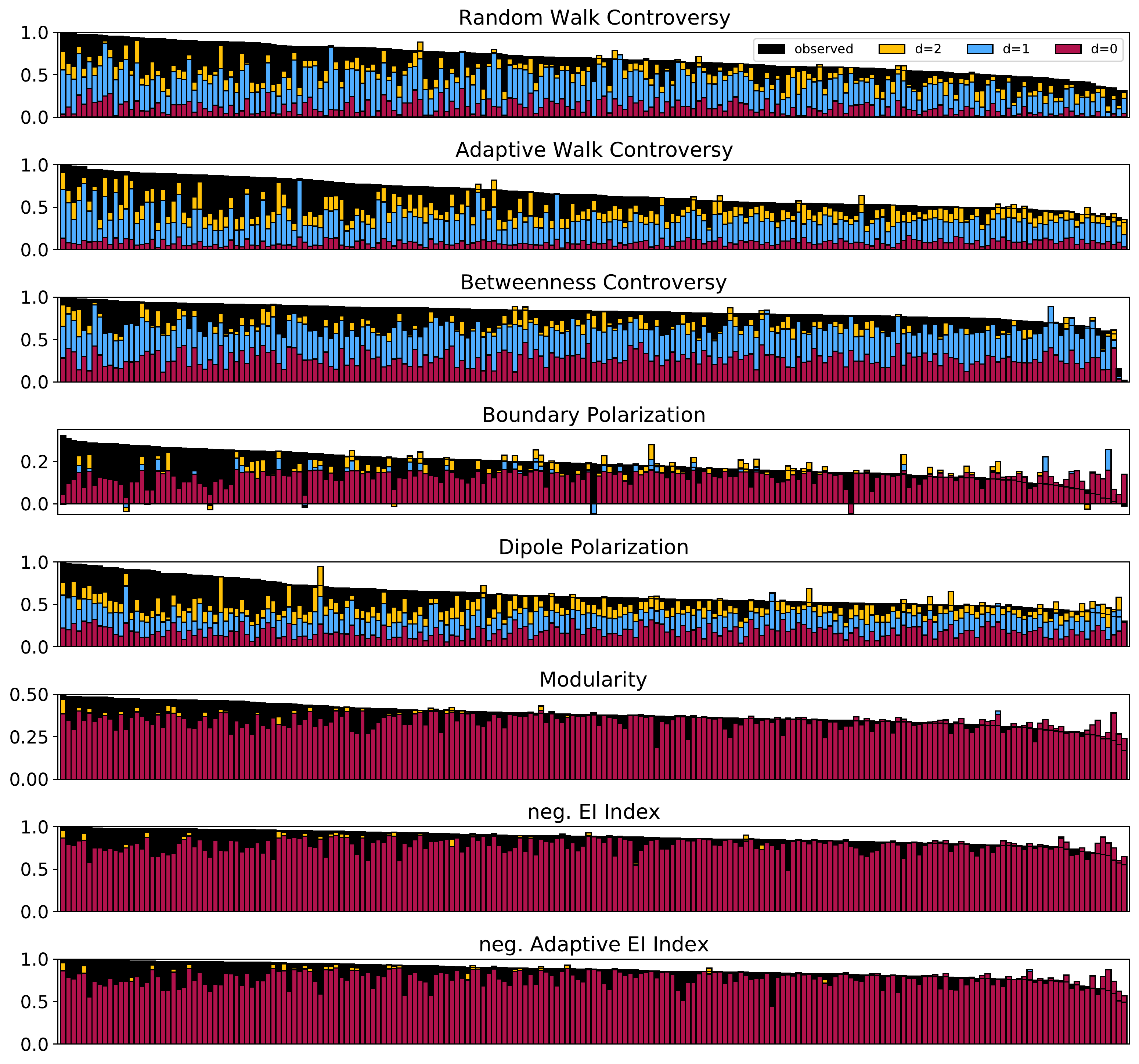}
    \caption{
    Polarization scores for the 203 observed networks and their shuffled versions. Each bar corresponds a score, and scores for a network and its randomized versions are on top of each other, ordered from bottom to top in the following order: observed network (black) and randomized networks where degree-degree sequence ($d=2$, yellow), degree sequence ($d=1$, blue), or average degree ($d=0$, red) is preserved. An interpretation for the figure is that, the amount of color that is shown tells how much of the total bar height (the score value) is explained by the corresponding network feature. Note that in some cases, the randomized networks produce higher scores than the original network and in this case the black bar is fully covered by the colored bar(s). In this case we draw a black horizontal line on top of the colored bars indicating the height of the black bar. See Appendix \ref{C} for similar results obtained by other partition algorithms.
    }
    \label{fig:bar}
\end{figure}

We first compare the observed network data to random networks that are shuffled in a way that we keep some of the features of the networks. As expected, the more features we keep, the more similar the scores are to the ones for original networks (see Fig. \ref{fig:bar}). 

For BP, Q, EI, and AEI the scores produced by the random models cover the observed score for most networks (black bar that corresponds to the observed score of a single network is covered by the other colors that correspond to the scores produced by different random models), indicating that the number of links and size of the networks ($d=0$) are already enough to predict much of the observed score. For RWC, ARWC, BC, and DP, the more features are kept, the higher (and therfore closer to the original value) the scores tends to be. In general, the change in scores after randomization follows a pattern where both low and high original scores can get very low values for the model keeping the average degree ($d=0$). The degree sequence ($d=1$) and degree-degree sequence ($d=2$) can in many cases explain most of the observed scores, and in some cases the scores for these random networks are even larger than for the original networks. We also have included an alternative way to visualize the polarization values of the randomized networks in Appendix \ref{C}. 

Fig. \ref{fig:bar} gives a detailed view of the poplarization scores. It can be used to read scores for each of the original networks and the corresponding random networks. There are four methods for which the $d=0$ model already explains most of the observed scores, and for the rest the degree sequence ($d=1$) is usually a very good predictor. 

\begin{figure}[ht]
    \centering
    \includegraphics[width=1\textwidth]{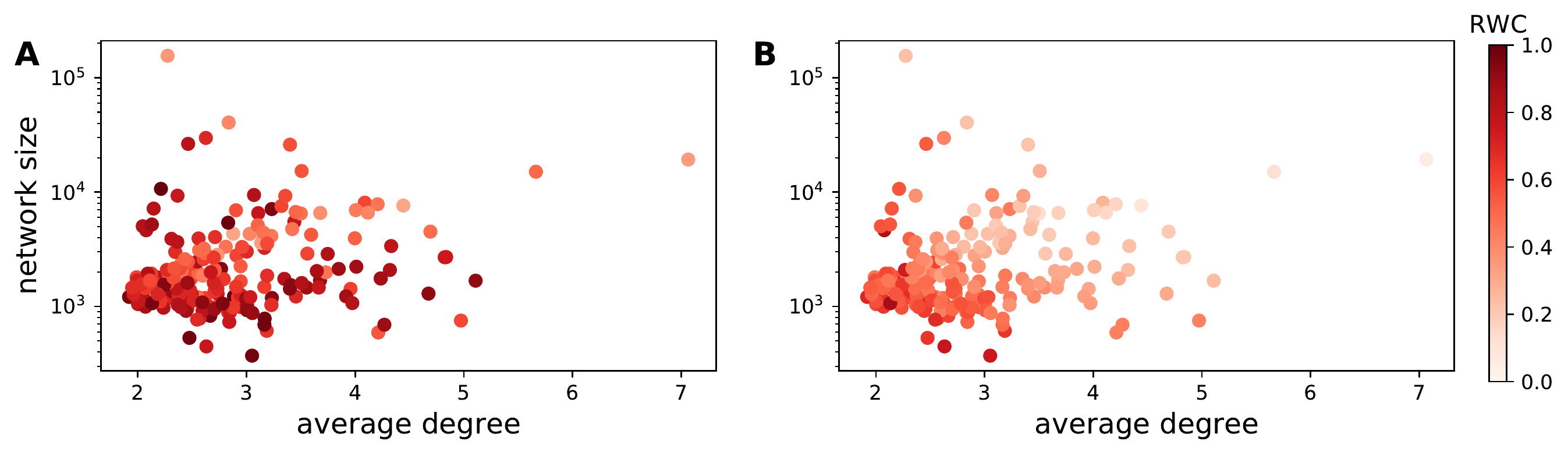}
    \caption{{\bf A} The RWC-scores for the 203 observed networks as a function of average degree and network size. {\bf B} The same score for configuration model $d=1$ of the real network. 
    }
    \label{fig:SCATTER}
\end{figure}


To get a more detailed view on the characteristics that are related with the polarization scores, we show the RWC score as a function of both network size and average degree in Fig. \ref{fig:SCATTER}. Network size is correlated in a way that smaller networks have higher RWC scores (Spearman correlation coefficient -0.42). 
After shuffling the real network with fixing the original degree sequence, the smaller and sparser networks have higher RWC scores (respective Spearman correlation coefficients -0.67 and -0.68).
Although randomized networks had lower RWC scores than the original networks, the averaged RWC for all networks with average degree less than four was approximately 0.45 even after randomization.


\subsection{Synthetic Random Networks}
Based on Section \ref{rnetana}, we see that polarization scores are heavily affected by elementary network statistics like network size, density, and degree distribution. We will next explore more systematically which factors explain the high polarization scores in randomized networks. In addition, to the aforementioned statistics, we analyse the effect of unequal community sizes as real network are likely to have a range of polarized group sizes.

\subsubsection{Network size and density.}\label{densitysection}

\begin{figure}[t]
    \centering
    \includegraphics[width=1\textwidth]{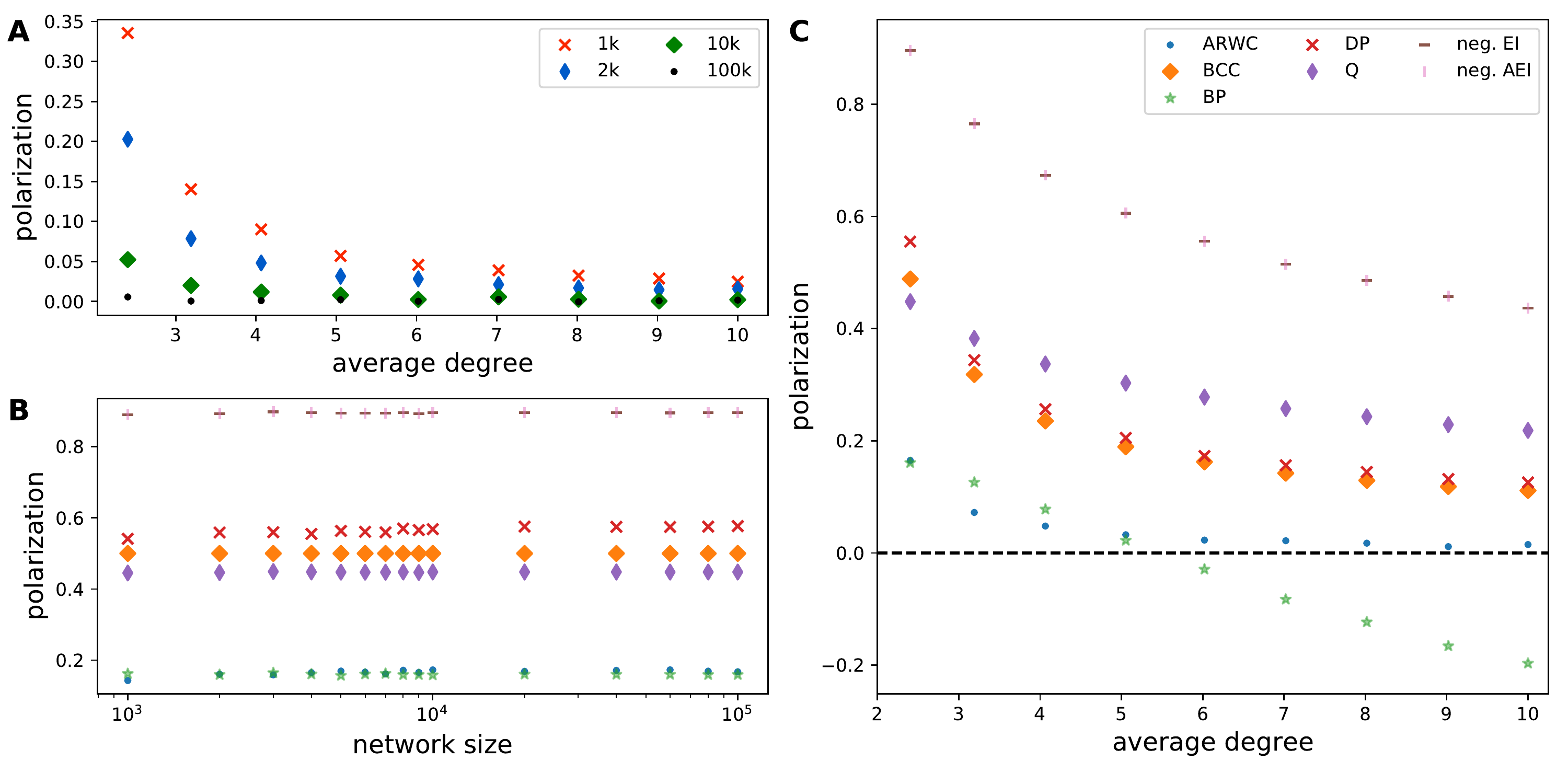}
    \caption{
    {\bf A} The expected RWC values as a function of average degree for ER-networks. The symbols denote the network size as indicated in the legend. {\bf B} The expected values of all polarization scores (except RWC) as a function of size for ER networks with average degree 2.4. See panel {\bf C} for legend. {\bf C} The expected values of all polarization scores (except RWC) as a function of average degree for ER networks with  4000 nodes. The dashed line indicates zero polarization.
    See Appendix \ref{C} for the polarization scores as a function of both number of nodes and average degree.
    }
    \label{fig:DEGSUM}
\end{figure}

Even an ideal structural polarization score can be correlated with network size and average degree in a specific set of real-world networks, but it should not be biased by these elementary network features in the set of all networks. 
As a consequence, structural polarization scores should get neutral values for random networks with any size and average degree that are generated without any explicit group structure that is present in the creation process.
To assess how the scores perform on this metric, we computed each score for Erdős–Rényi graphs of varying sizes and densities. Our results shown in Fig. \ref{fig:DEGSUM} indicate that all scores are affected by at least one of these elementary network features.

First, we find that network size generally did not contribute to polarization scores, with RWC being the sole exception. It was affected by the number of nodes in the network, giving lower values for larger graphs with the same average degree (see Fig. \ref{fig:DEGSUM}). For networks with enough nodes, the RWC is very close to zero for all values of average degree, but this comes at the cost of the score being sensitive to network size. On the other hand, despite being a similar random walk-based score, the ARWC is invariant to network size. This highlights the difference in their construction. Specifically, the RWC takes as a parameter a fixed number of influencers in the system, meaning that the number of influencers \textit{as a proportion of the network} varies by network size, leading to inconsistent variation in RWC across networks. The ARWC removes this dependence by setting the number of influencers as a function of network size (i.e., as a proportion of network size it remains constant). We discuss the difference between the RWC and ARWC in Appendix \ref{A}.

Specific instances of RWC aside, all scores are dependent on average degree, and only approach non-polarized levels when the network's average degree exceeds a certain value. For instance, BCC gives zero polarization for random graphs only when the average degree is approximately 10. This is quite strict condition to have especially for Twitter networks. BP decreases almost linearly as a function of density. It reaches the value zero when the average degree of the network is between 5 and 6. The negative values for larger degrees indicate that nodes in the boundary are more likely to connect to  the other group. Morales et al., i.e., author behind the DP score, pointed out how their score suffers from the ``minimum polarization problem'' due to the nonzero standard deviations of the opinion distributions \cite{morales2015measuring}.

Dependence between density and modularity has been studied theoretically before for the case where the number of clusters is not restricted to two like in polarization scores. 
Previous research has shown that,
sparse random graphs (and scale-free networks) can have very high modularity scores \cite{guimera2004modularity}, and that, with high probability, modularity cannot be zero if the average degree is bounded \cite{mcdiarmid2020modularity}. 
It is therefore known that using a network's modularity score to measure the amount to which it is clustered is inadvisable.
This notion has been speculated to apply for the use of modularity as a polarization score \cite{guerra2013measure}.
Our results confirm this notion, and show that modularity behaves similarly for the case where the number of clusters is limited to two, with the difference that the maximum value in our simulations goes up to only approximately 0.5.
Here, it should be noted that none of the other scores seem to be immune to this problem. 

\subsubsection{Heterogeneity of degree sequence.}

\begin{figure}
    \centering
    \includegraphics[width=1\textwidth]{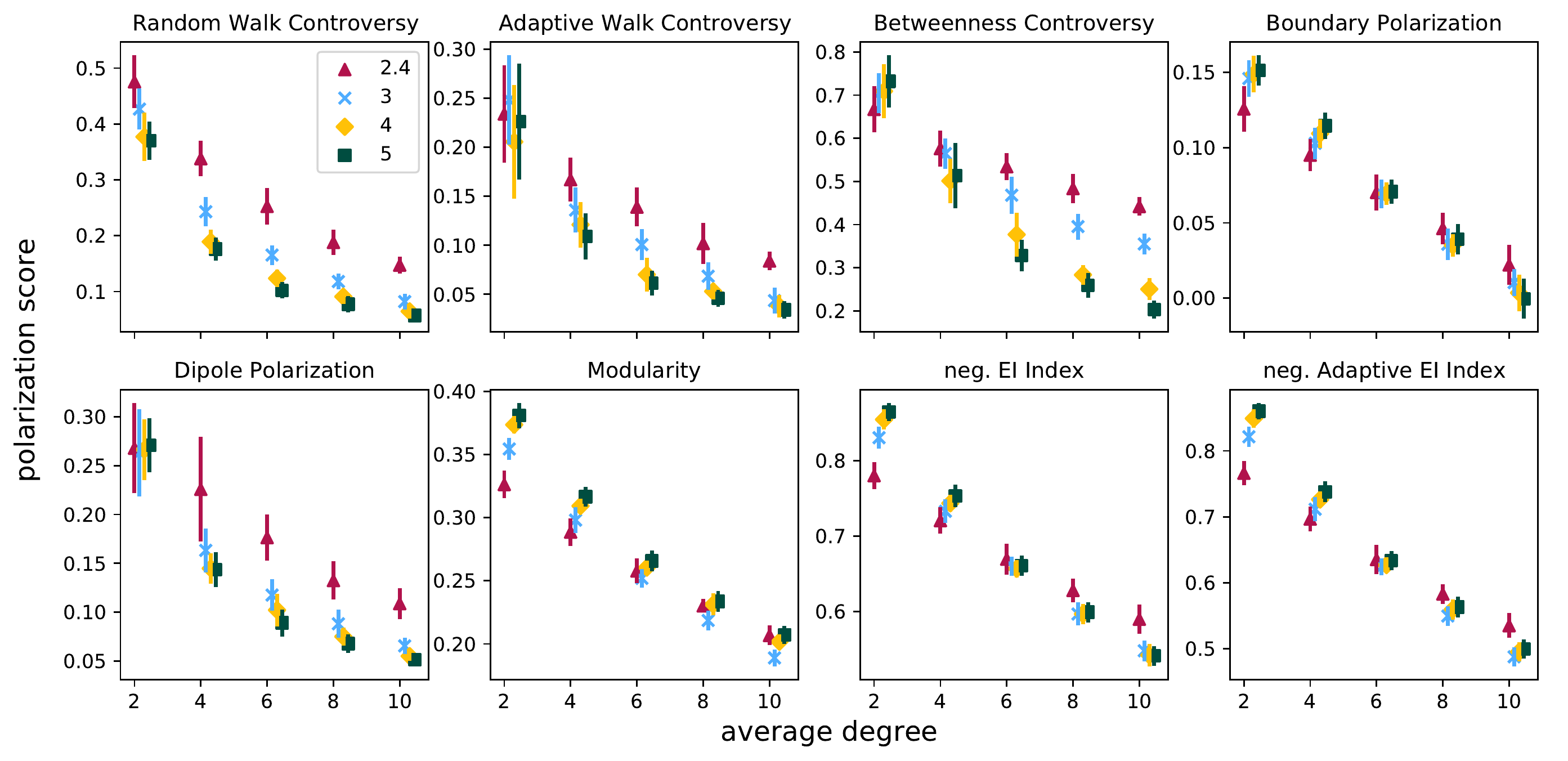}
    \caption{The effect of degree heterogeneity to polarization scores for the 8 scores in simulated scale-free networks with 1000 nodes. We show the expected polarization score and its standard deviation as a function of average degree. The symbols correspond to different exponent $\gamma$ values as indicated by the legend. See Figs. \ref{fig:deghet2k}-\ref{fig:deghet5k} in Appendix \ref{C} for similar results for larger networks.
    }
    \label{fig:deghet1k}
\end{figure}

The role of the underlying degree sequence is essential to study as political communication networks tend to have a few nodes with relatively high number of edges.
For networks produced by the Erdős–Rényi model, the degree distribution is Binomial centered at the average degree $\langle k \rangle = p(n-1)$, where $p$ is the probability that two nodes are connected and $n$ is the number of nodes in network. 
In contrast, many real networks' degrees follow fat-tailed distributions which can have considerably higher variance \cite{bara,clauset2009power,serafino2021true}.
To analyze the effect of degree distribution,
we simulate random graphs whose degree sequences were drawn from a power law distribution $P(k) \propto k^{-\gamma}$ \cite{gene}. 
We vary the exponent $\gamma$, which allows us to explore the amount of variation in the degree distributions. The lower the value of $\gamma$ is, the higher is the variation and larger are the hubs.

The RWC, ARWC, BCC, and DP give higher scores for more heterogeneous networks, and there is a slight opposite effect for the other scores given a low average degree (see Fig.~\ref{fig:deghet1k}).
The average degree affects polarization scores for networks with heterogeneous degree sequences as well.
The observation that the level of polarization approaches zero only when the network becomes denser still holds for all scores, but for the scores that are lifted by degree heterogeneity, much denser networks are required in order for the scores to become close to zero.

\subsubsection{Communities of different sizes.}\label{sbmsection}

\begin{figure}
    \centering
    \includegraphics[width=1\textwidth]{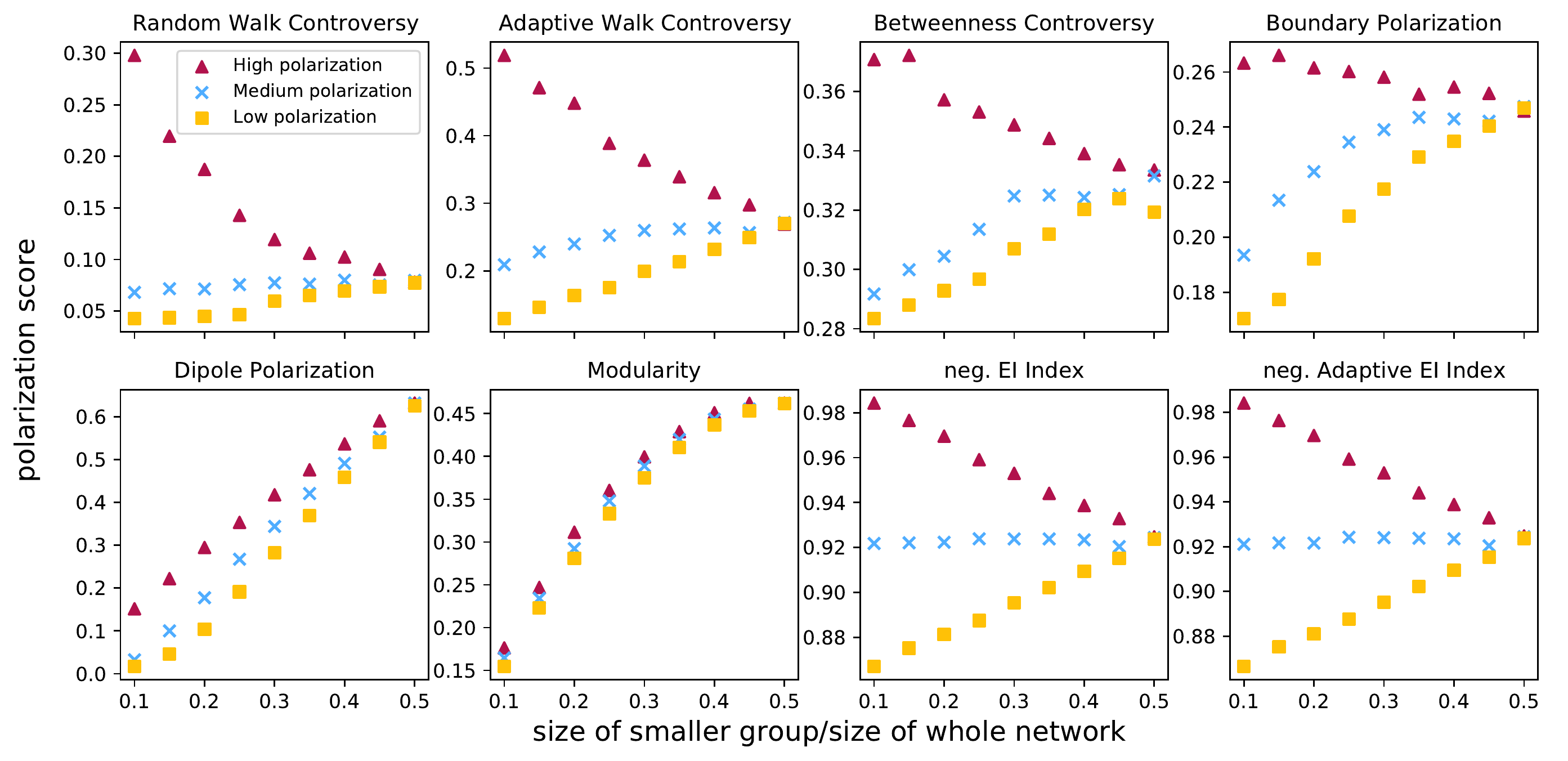}
    \caption{Polarization scores as the function of relative group size $n_{\text{small}}/(n_{\text{small}} +n_{\text{large}})$ in an SBM model described in the main text. The different markers correspond to the three schemes of adding inter-block links. 
    }
    \label{fig:comsize}
\end{figure}

Polarized groups within a system are often imbalanced in size, making it important for scores to perform consistently across a range of group size balance. To assess this metric, we used the stochastic block model to generate a set of polarized networks differing by group size imbalance and level of polarization. All networks were fixed to have size $n=10000$, and the size of the smaller group ranges from between 10\% to 50\% of the whole network. Within-group links were generated for each node to have an average within-group degree of $k_{\text{in}}$. To obtain different levels of polarization, we generated between-group links at three levels of density. For \textit{low polarization} networks, links were generated with the expectation of adding $k_{\text{out}}=k_{\text{in}}/c$ between-group links per node from the larger group. Conversely, \textit{high polarization} networks have an expected $k_{\text{out}}$ between-group links per node from the smaller group. A third set of networks, with \textit{medium polarization}, have an expected $k_{\text{out}} \times n /2$ total between-group links. Our networks were generated with $k_{\text{in}}=4.5$ and $c = 25$, but our results are robust to a reasonable range of different values.


Our results indicate that all scores depend on group size imbalance at least for the high and low polarization schemes (see Fig. \ref{fig:comsize}). The EI and AEI scores are relatively insensitive to the group size imbalances as their values change by only a few percentage points. For all scores except DP and Q, simulated level of polarization affects the extent to which they are dependent on group size imbalances; at certain levels of simulated polarization, the dependence disappears. For EI and AEI this level is exactly the one present in the medium polarization networks.
Finally, it is worth noting that DP has an intrinsic scaling factor which penalizes the difference between groups. Specifically, its coefficient is designed to contribute maximally to the polarization only when the communities are equally-sized, thus the linear relationship between imbalance and score.




\section{Evaluation and Normalization of the Polarization Scores}
\label{sec:evaluation}

As our analysis suggests, nonzero polarization values arise in randomized networks due to the scores being sensitive to network size, number of links, and degree distribution. These features in themselves are not likely to be linked to polarization. The number of nodes and the number of links depend on unrelated system features such as overall activity in the network and and design choices such as link sampling schemes. Further, the fat-tailed degree distributions we often observe in the data have been attributed to preferential-attachment mechanisms \cite{bara}. 

However, even if scores do not return zero values for random networks or are biased by some features of the data, they can still be useful if they are able to separate polarized networks from non-polarized ones by systematically returning higher scores for the former set. In this section, we assess the predictive validity of the scores against a manually labeled set of 203 polarized and non-polarized networks, which we introduced in Section \ref{datasection}. The 20 networks from Garimella et al.'s study are already manually labeled \citep{garimella2018quantifying}, so we are able to directly use these external labels. The 183 networks from Chen et al.'s study had not been labeled \citep{chen2020polarization}, so we manually labeled them for our present task. 

We labeled these networks based on the content of tweets, which ensures that our labeling is independent of the structural polarization measures. We checked whether any of the pre-defined features, which are selected based on our definition of polarization, were present in a reasonable number of tweets before marking the network polarized. The labeling process, which we describe below, was performed before the main analysis, and resulted in a balanced data set with around $47\%$ of the networks labeld as polarized. To the extent that the basic network features we examined are not indicative of polarization, we should see an increase in the classification performance of the scores when the effects induced by them are removed.

We recognize that polarization labels based on content alone are not necessarily the ground truth on polarization in the system. Instead, content-based labeling is another method for quantifying polarization, which is based on a set of criteria different from those used in structural polarization scores. Because it is another label of the same underlying latent construct, content-based labeling is useful for assessing the validity of structural polarization scores. If a structural score correlates well with our content-based labels, they are said to have high \textit{convergent validity}, which is a form of measurement validity\cite{adcock2001measurement}. This means that it can be seen as a better measure of the latent overall polarization in the system compared to a less correlated structural score. 



\subsection{Data Labeling}\label{B}
We labeled our networks before performing the main analysis. All networks with at least one hashtag containing the substring `election' were classified as polarized. We manually sampled tweets from each network for confirmation. For each network, we applied a four-stage process for labeling in the following order:

\begin{enumerate}
  \item Sample uniformly 5 days from which tweets are read.
  \item Display all the tweets from each sampled day.
  \item Sample 20 users from each sampled day and display all their tweets during the sampled days.
  \item Partition the network and distinguish 10 highest degree nodes from both groups. Display all their tweets.
\end{enumerate}
After displaying the tweets that were obtained by the described process, we checked whether any of the following features were present in a reasonable number of tweets.

\begin{itemize}
    \item us-versus-them mentality, signs of disagreement, dispute, or friction
    \item strongly discrediting the out-group or strongly supporting the in-group from both sides
    \item direct, negative, or strong criticism of political adversaries or political actors from both sides
    \item completely opposite opinions, beliefs, or points of view on a political or social topic
\end{itemize}
Based on the content of the sampled tweets together with domain knowledge, a researcher classified the network either polarized or non-polarized. If the sample was too vague to be labeled, we repeated the process to gain a clearer view into the general content of tweets.

\subsection{Classification Performance of the Polarization Scores}
\begin{figure}[t]
    \centering
    \includegraphics[width=1\textwidth]{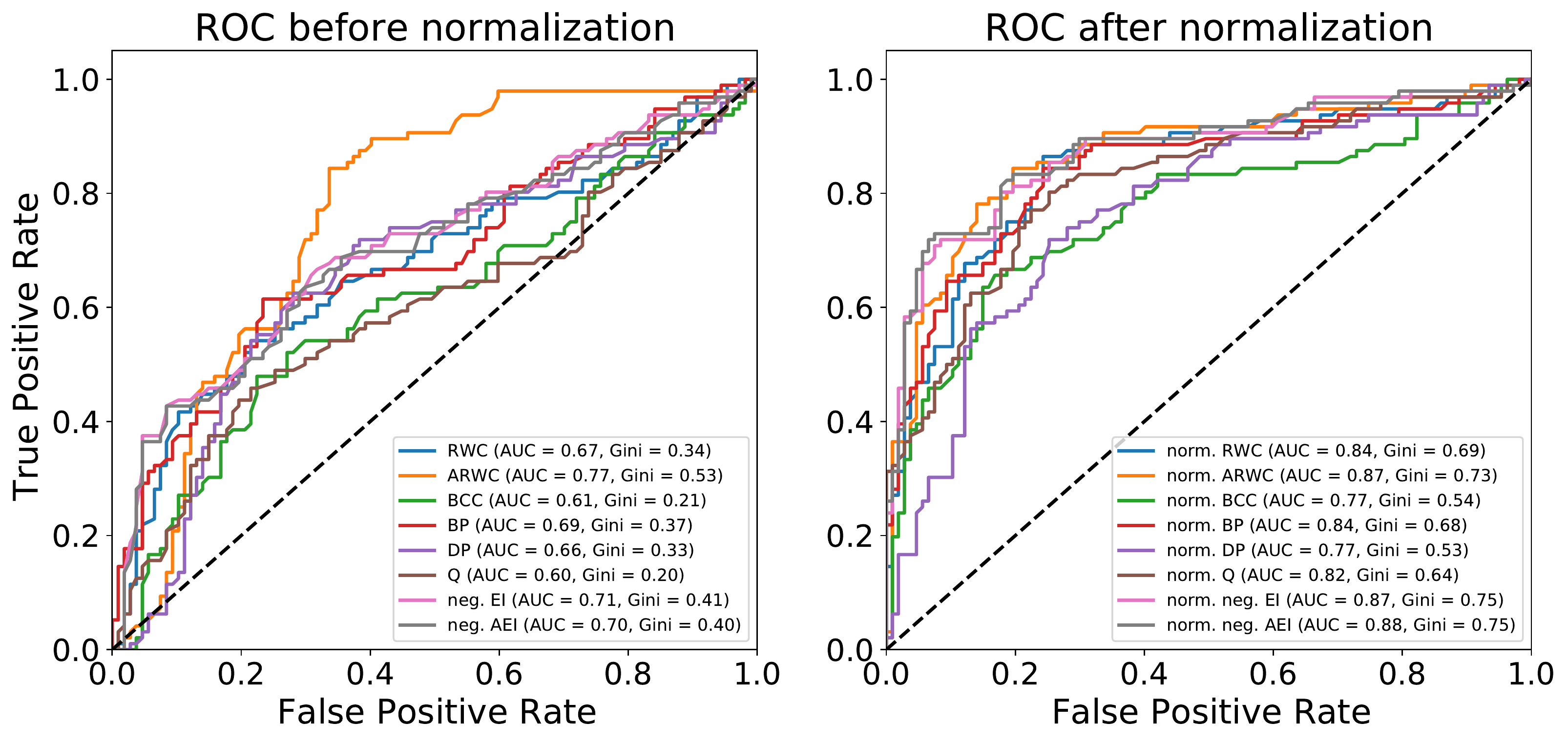}
    \caption{ROC curves, Gini coefficient values, and AUC values for the task of predicting manually curated labeling of polarized and non-polarized networks. The results shown (left) for the score values before the normalization and (right) after the normalization with denoised scores $\hat{\Phi}(G)$  (see main text).
Figures are based on all the 203 empirical networks described in section \ref{datasection}. Results for the different normalization scheme $\hat{\Phi}_z(G)$ including the standardization are shown in the Fig. \ref{fig:ROCXX} in Appendix \ref{C}. The ROC curves for the alternative graph partitioning methods are also reported in Figs. \ref{fig:ROC_SPECTRAL} and \ref{fig:ROC_MOD} in Appendix \ref{C}.}
    \label{fig:roc}
\end{figure}

We adapt a typical framework where there is a decision threshold for the score value under which all networks are predicted to be non-polarized and above which they are prediced to be polarized. Each threshold value produces a false positive rate and true positive rate, and varying the threshold gives us an ROC curve (shown in Fig. \ref{fig:roc}) characterizing the overall performance of the scores in the prediction task. This makes the evaluation independent on the selected type of classifier. We also derive single scores to quantify the overall performance. The \textit{Gini coefficient} measures the performance improvement compared to a uniformly random guess and is defined as the area between the diagonal line and the ROC curve, normalized so $1$ indicates perfect prediction and $0$ indicates random guess. We also report the unnormalized area under curve (AUC).

The Gini coefficients for the scores vary between $0.20$ and $0.53$ with Q and BCC performing the worst. ARWC performs the best with a wide margin to the second best AEI and EI (with coefficient values $0.40$ and $0.41$). The non-adaptive RWC has a Gini coefficient of $0.34$, which is better than prior work shows \cite{emamgholizadeh2020framework}, but still generally poor. Notably, the ARWC score performs very well if we accept a high false positive rate. That is, if a system has a very small ARWC score, it is a good indication of the network not being polarized. In contrast, a large ARWC score (or any other measure) does not necessarily indicate that the network is polarized. As our prior results show, a high score might be due to the effects of small and sparse networks with hubs as opposed to anything related to polarized group structures.

\begin{figure}[t]
    \centering
    \includegraphics[width=1\textwidth]{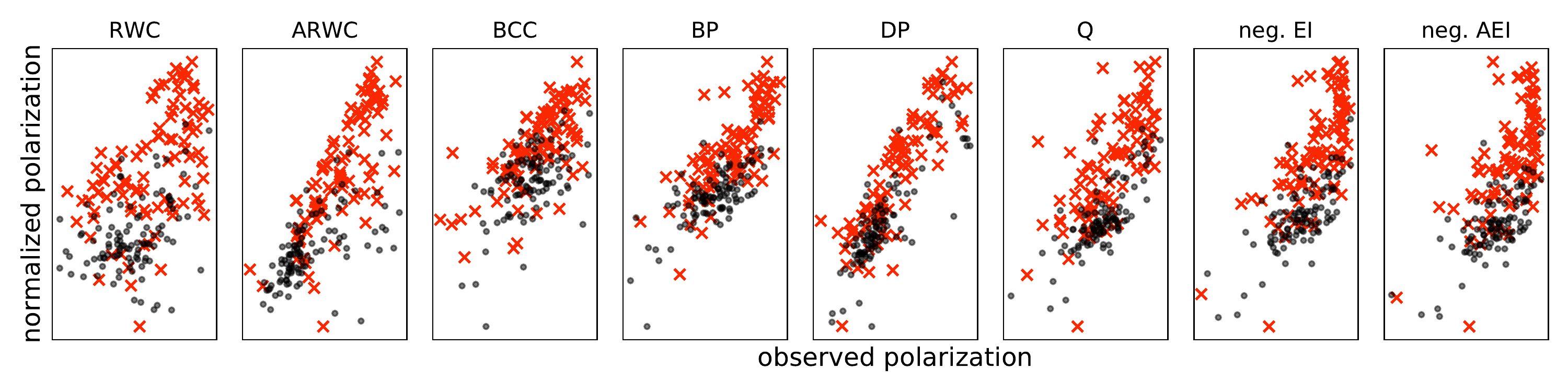}
    \caption{
    Scatter plot of denoised ($\hat{\Phi}$) and observed polarization scores for the 203 networks described in Section \ref{datasection}.
    Red crosses are networks labeled as polarized and black points are networks labeled as non-polarized. An outlier was removed from the plots. 
    Note that the scales for the scores are different and not shown in this figure. See Fig. \ref{fig:scatterstandard} in Appendix \ref{C} for the equivalent scatter plot for the standardized scores ($\hat{\Phi}_z$).
    }
    \label{fig:normpanel}
\end{figure}

\subsection{Normalization of the scores}\label{normsection}



To remove the effect of network size and the degree distribution, we computed the averaged polarization score for multiple instances of the network shuffled with the configuration model ($d=1$ in the $dk$-framework), and  subtracted it from the observed polarization score value. That is, given a network $G$ and a polarization score $\Phi$, we define a normalized score as
\begin{displaymath}
\hat{\Phi}(G) = \Phi(G) - \langle\Phi(G_{CM})\rangle\,,
\end{displaymath}
where $\Phi(G)$ is the polarization score of observed network and $\langle\Phi(G_{CM})\rangle$ is the expected polarization score of graphs generated by the configuration model. This score corrects for the expected effect of the size and degree distribution of the network (i.e. removes the blue part from the observed score shown previously in Fig. \ref{fig:bar}). Thus we call it \textit{the denoised polarization score}. It does not consider the fact that there is some fluctuations in the score values in the configuration model. We correct for this in another, slightly different normalization scheme, where we divide the normalized score by the standard deviation of the score value distribution for the configuration model:
\begin{displaymath}
\hat{\Phi}_{z}(G) = \frac{\Phi(G) - \langle\Phi(G_{CM})\rangle}{\sqrt{\langle\Phi(G_{CM})^2\rangle - \langle\Phi(G_{CM})\rangle^2}}\,.
\end{displaymath}
We call this normalization \textit{standardized and denoised polarization score}. Note that the distribution of polarization value over the ensemble of null random graphs is not necessarily Gaussian. If the values $\Phi(G_{CM})$ followed Gaussian distribution, then the statistical significance testing could be performed with the standard normal distribution, and $\hat{\Phi}_{z}(G_{CM})$ would be the appropriate test statistic (the z-score). An approximate value for a significance can be obtained with large number of samples. The same normalization has been proposed for modularity to mitigate the resolution limit \cite{miyauchi2016z}. In that work, the proposed quality function gives higher values for statistically rarer observations.


Fig. \ref{fig:roc} shows the ROC curves and related coefficient values for the denoised  scores $\hat{\Phi} (G)$ (see the qualitatively similar results in Appendix \ref{C} for $\hat{\Phi}_z(G)$). The performance of all of the scores increased after normalization as indicated by the overall lifted ROC curves and improved Gini coefficients (and AUC values). Improvements in Gini coefficients after normalization ranges between 38\% - 220\% depending on the measure used. The ARWC remains among the best performing scores, along with the AEI and EI. The AEI in particular, perfoms best under conservative estimates of polarization (i.e., low false positive rates). The BCC is still among the worst after the normalization, along with DP. However, all post-normalization scores notably outperform the best unnormalized score (ARWC).



Fig. \ref{fig:normpanel} further illustrates the dramatic change in the predictive power of the scores after normalization. It shows the normalized score values as a function of the observed score values. With most scores, the two types of networks are mixed together when sorted with the observed scores (on the x-axis). Normalization largely lifts the polarized networks above the cloud of non-polarized ones, making it possible to separate these two groups (on the y-axis).  
This finding holds for all the polarization measures analyzed here. 

Note that in practice, to implement the normalization procedure, one needs to randomize the network multiple times with the configuration model and compute the score $\Phi(G_{CM})$ to get estimates for the mean value of the score $\langle\Phi(G_{CM})\rangle$ (and $\langle\Phi(G_{CM})^2\rangle$ for $\hat{\Phi}_{z}(G)$). Here we sampled the networks 500 times which was more than enough samples as they lead to error of the means ranging from 0.01 to 0.05.

\begin{figure}[tb]
    \centering
    \includegraphics[width=1\textwidth]{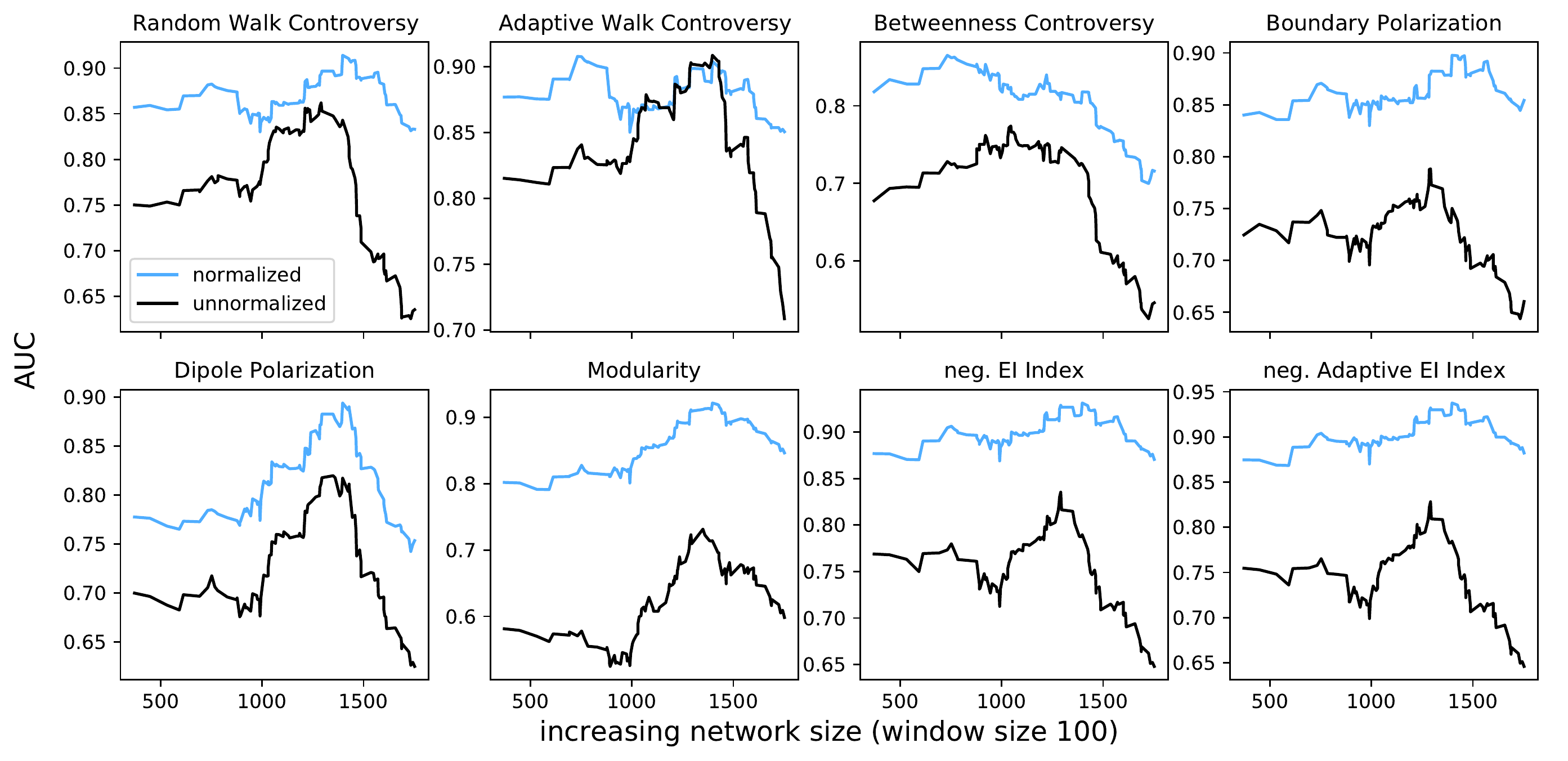}
    \caption{
    Quantifying how network's average degree affects the performance. We group the data such that there are 100 networks with consecutive sizes in our data, and create a set of such windows by varying the size range. We then evaluate the AUC for the moving window of 100 networks.
    Generally, all the networks benefits from the normalization across all the polarization methods. The same analysis for average degree is included in Appendix \ref{C}. Also the details on the scale of the windows are included in Appendix \ref{C}.}
    \label{fig:new_aucsize}
\end{figure}

To determine which types of networks benefits the most from the normalization, we plotted the AUC as a function of network size and average degree for each polarization score. This was done by evaluating the performance for subsets of networks with a fixed window size of 100 (shown in Fig. \ref{fig:new_aucsize}). The results show how the performances for all the polarization methods are better and more stable after the normalization independent on the network size. Only ARWC has a short region where the performances of both normalized and unnormalized scores overlap. The same analysis for average degree is included in Appendix \ref{C}. 

We also tested whether combining the results from all polarization methods improves the accuracy of predicting whether a network is polarized. Our first strategy was to take the average of all the scores and use that as a new polarization score. The second strategy was to train a simple decision tree classifier where the input vector contained all eight scores obtained for a network. 
The AUC for the average of unnormalized values was 0.71, and for normalized values it increased to 0.87. Although the averaged normalized score outperformed some of the single normalized polarization scores (e.g. BP, DP, and Q), it did not outperform the best-performing ones (e.g. ARWC and AEI). Regarding the decision tree, the AUC of the pre-normalization classifier was 0.78, whereas for the post-normalization one, the AUC increased to 0.90. Our results show that strategies based on combined scores can in some cases offer improvements over single polarization scores, but only minimally. It is up to the researcher to decide if these gains are worth the cost of additional work and loss of transparency associated with training machine learning models. 

\section{Discussion}
Measuring polarization is important for social science research, including the social computing and computer mediated communication fields. Structural polarization measures offer an ostensibly promising approach, but we identified a number of undesirable properties associated with all eight commonly-used measures studied here. These measures can be high for random networks and they are sensitive to various network features such as size, average degree, and degree distribution. These properties pose a clear problem for polarization research. Considerable research effort has been put into polarization identification and platform design for attenuating polarization, but if the measurement of polarization is systematically biased by basic network features, our ability to make valid inferences are greatly reduced. For example, consider Bail et al.'s study that found increasing exposure to opposing views increased political polarization \citep{bail2018exposure}. The study did not rely on structural polarization measures, but had this study been conducted in the field using the RWC to measure polarization, the increased activity that likely would have resulted from the intervention could have decreased polarization scores, resulting in the exact opposite conclusion being drawn.

Based on our results, we strongly recommend applying the normalization procedures introduced in Section \ref{normsection} for applied work using any of the network-based polarization scores included here. Doing so removes a substantial amount of noise arising from the network's local properties. For our test networks, classification performance improved by 38\%-220\% (Gini coefficient) depending on the measure. Notably, the differences in performance across polarization scores were minor after normalization. In fact, the AEI and EI, which are the simplest and least computationally-demanding scores, were among the best performing scores.


In order for us to draw qualitative conclusions based on the score values we should understand the scale of the score, e.g., what values constitute medium or high polarization. The unnormalized score values often have interpretation described in the logic of their definition.
Despite their relatively high accuracy, normalized scores are less amenable to this kind of direct interpretation. If this is needed, a plausible alternative is to report both the score itself and its expected value in one or more random network models. This way, one has a sense of how much of the score is explained by the various features of the network. 

Our work has implications for additional structural polarization scores not studied here, including those in development. It is clear from our results that structural scores obtained via the consecutive procedures of network clustering and score computation are susceptible to being sensitive to various network features in a way that is not apparent from the score's definition. Our argument (and others' before us \cite{guerra2013measure}), backed up by the results that normalization increases the performance of the scores, is that these sensitivities bias the scores. At a minimum, one should have a clear idea of how the given score behaves in relation to basic network statistics and group sizes. 
To facilitate such examination, we have made our analysis pipeline and data publicly available \cite{code1,code2}.
There could be other possible sources of bias, so our benchmarking framework should be taken as a minimal test that is not necessarily sufficient.

More broadly, the fact that all eight scores we tested were affected to some extent by the same problems suggests that the approach of separating polarization scoring into independent clustering and score calculation phases might be flawed. This is part of the wider problem where clustering methods start from the assumption that the given network contains clusters and can find them even in random data. A solution to this problem could be to break the independence of the clustering phase and the score calculation phase, using instead clustering methods that can test if the clusters could be explained by random networks \cite{lancichinetti2010statistical,peixoto2013parsimonious}. Scores can be set to zero if no significant clusters are found. This reduces the false positive rate, which was especially problematic with the best-performing ARWC method.

Our study presents some limitations which can be avenues for future research. First, our results are based purely on simulations, and a more theoretical approach could be taken to understand the individual scores better. This work can build on the prior work on fundamental problems in graph clustering which, as illustrated here, are reflecting onto polarization scores. In this context, modularity is a well-studied example of how an apparently reasonably defined method can have many underlying problems \cite{guimera2004modularity,fortunato2007resolution,bagrow2012communities,mcdiarmid2016modularity,mcdiarmid2020modularity}.
Given this, analyzing modularity from the perspective of limiting the number of clusters to two could be done as an extension to the previous literature on modularity optimisation with an arbitrary number of communties. Even if modularity is not important as a structural polarization score, this analysis could shed light on the kind of phenomena to expect when scoring networks with two clusters. 

Second, public opinion on politics can be multisided. This means that instead of having only two groups, there can be multiple cohesive clusters that are segragated in some way in the network. However, the majority of polarization measures, including the structural measures analyzed here are defined exclusively for two cluster, with the exception of modularity and the EI-index. Conceptual and technical work that generalizes polarization to the multiside context is therefore useful. This is a nascent area of study \cite{reiljan2020fear}, with some extensions to structural measures \cite{markgraf2019quantification,gaumont2018reconstruction}.
Such generalizations are likely to retain the same problems as their two-sided variants, because more degrees of freedom in the number of groups for the clustering algorithms will lead to better clusters (as measured with the internal evaluation metrics of the methods). 
As discussed in section \ref{densitysection}, previous work on modularity can again be useful here, as it indicates that the issues high score values in random networks is even worse when the number of clusters is not limited.

Further, a clear limitation of the current work is the number and variety of labeled network data that was used. While the number of network is enough to statistically show that normalization improves the score performance, a more fine-grained view of the problem could be achieved with more networks. Similarly, the generalizability of the classification results could be improved by widening the range of network size and density but more importantly by including different types of social networks.
Here, it is worth noting that our approach to labeling the content might not be as clear cut for other contexts such as non-political communication.
Finally, the analysis regarding the different-sized clusters can be improved. Although our results indicated that all scores depend on group size imbalance at least for the low and high polarization schemes, other techniques for simulating polarization between the communities should be examined. 

Despite the issues we raised in this paper, structural polarization measures as an approach remains useful. In addition to being based on emergent behavior directly observed in the system under study, they facilitate an accessible approach to studying polarization. A network-based approach generally has low language processing requirements, making it equally easy to apply to different linguistic contexts. Additionally, it has been argued that there is an uneven development of natural language processing tools across languages \citep{djatmiko2019review}, which presents an additional barrier to content-based polarization measures. Content-based polarization measures often require language-specific resources such as sentiment lexicons, which are often costly to build \citep{sun2017review}. Similarly, survey-based measures, especially from original data sources, are costly to obtain. In light of these sometimes considerable barriers to research, structural polarization measures provide an accessile alternative for applied researchers.

To facilitate the use of structural polarization measures, we introduced in this paper a minimal set of tests that a structural polarization score should pass in order for it to be able to distinguish polarization from noise. This should serve as a benchmark for future developments of such scores. The fact that the all of the current scores perform relatively poorly indicates that there is a need for alternative approaches to the typical scoring approach. The normalization procedure we introduced here is a patch that alleviates this problem. There is space for other, possibly fundamentally different approaches, to be innovated for measuring structural polarization.

\begin{acks}

This research is part of the ECANET-consortium (Echo Chambers, Experts and Activists: Networks of Mediated Political Communication), part of the Media and Society research programme (2019-2022) funded by the Academy of Finland (grant number: 320781).
\end{acks}

\bibliographystyle{ACM-Reference-Format}
\bibliography{sample-base}

\appendix

\section{Polarization measures}\label{A}

Here we briefly introduce the definitions of polarization measures studied in the main body of the paper. Each score here assumes two disjoint sets $A$ and $B$. Here a network refers to the giant component only. Let $V=A\cup B$ be the set of nodes and $E$ be the set of edges in network. The membership $c_i$ of the nodes are determined during the network partition stage. In a polarized network, the sets are expected to represent the opposing communities or sides. Some of the scores, such as Random Walk Controversy and Boundary Polarzation, are also designed to capture potential antipolarization behavior.

\begin{enumerate}

  \item Random Walk Controversy \newline
  This measure captures the intuition of how likely a random user on either side is to be exposed to dominant content produced by influencers from the opposing side. From both sets $k$ nodes with the highest degrees are selected and labeled as influencers. The high-degreeness of a node is assumed to indicate a large number of received endorsements on the specific topic, thus called influencers.
  
  A random walk begins randomly from either side with equal probability and terminates only when it arrives in any influencer node (absorbing state). Based on the distribution of starting and ending sides of the multiple random walks, the score is computed as 
  
  $$P_{RWC} = p_{AA}p_{BB} - p_{AB}p_{BA},$$
  
  where $p_{AB}$ is the conditional probability for a random walk ending in side $B$ given that it had started from side $A$. The other probabilities in the formula are computed similarly. The polarization score $P_{RWC}$ takes values between -1 and 1. Fully polarized network has a $P_{RWC}$ of 1, whereas a non-polarized network is expected to have $P_{RWC} = 0$. If the users are exposed more likely to the content produced by influencers of the opposing group, the $P_{RWC}$ becomes negative.
  
  As there was no general network dependent rule for choosing the parameter $k$ in \cite{garimella2018quantifying}, we chose a single value $k=10$ for all of the networks.
  
  \medskip
  
  \item Adaptive Random Walk Controversy 

  The Random Walk Controversy measure is very sensitive to the number of influencers $k$. As no strategy for selecting the parameter $k$ based on the network was presented in the article it was defined \cite{garimella2018political},
  we devised such a strategy to adaptively change $k$ depending the network based on an initial sensitivity analysis of the score.
  Instead of selecting a fixed number of influencers from both sides, the number of influencers in a community depends on its size. By labeling the fraction $K$ of the highest-degree nodes as influencers from each side, i.e., by selecting $k_A=K/n_A$ for the community $A$ and $k_B=K/n_B$ for community $B$ with fixed $K$, the polarization measure $P_{ARWC}$ scales with the number of nodes in community. We used $K=0.01$.
  It should be noted that the actual values for the ARWC score (and RWC score) are sensitive to these parameter choices making comparison of results difficult if different values are used, but the qualitative behavior relative to random networks as described in the article is not sensitive to small changes in the actual parameter value $K$ (and $k$ for RWC). 
  \medskip
  
  \item Betweenness Centrality Controversy \newline
  This measure is based on the distribution of edge betweenness centralities. If the two sides are strongly separated, then the links on the boundary are expected to have high edge betweenness centralities. The intuition is that, in highly polarized network, links connecting the opposing communities have a critical role in the network topology.  The centrality $c_B$ for each edge present in the network is defined as follows
  
  $$c_B(e) = \sum_{s,t \in V}\frac{\sigma(s,t|e)}{\sigma(s,t)},$$
  
  where $\sigma(s,t)$ denotes the total number of shortest paths between nodes $s$ and $t$ and $\sigma(s,t|e)$ is the number of those paths that include edge $e$. 
  
 Then KL-divergence $d_{KL}$ is computed between the distribution of edge centralities for edges in the cut and the distribution of edge centralities for the rest of edges. The PDFs for KL are estimated by kernel density estimation. The measure seeks to quantify polarization by comparing the centralities of boundary and non-boundary links. The measure is defined as
 
  $$P_{BCC} = 1-e^{-d_{KL}},$$
  
  The score approaches 1 as the level of separation increases. For networks in which the centrality values of links between two communities do not differ significantly, $P_{BCC}$ produces values close to 0. 
  
  \medskip
  \item Boundary Polarization
  
  This measure assumes that a low concentration of high-degree nodes in the boundary of communities implies polarization. The underlying intuition is that the further some authoritative or influential user is from the boundary, the larger is the amount of antagonism present in the network. Two sets, $C$ and $I$, are defined for the score. The node $s\in A$ belongs to set $C_A$ if and only if it is linked to at least one node of the other side ($t\in B$) and it is linked to a node $w\in A$ that is not connected to any other node of side $B$. For the whole network, we have $C = C_A \cup C_B$, as both sides have their own boundary nodes naturally. The non-boundary nodes are called internal nodes and are obtained by $I_A = A-C_A$. The sets of internal nodes of both communities are then combined by $I = I_A \cup I_B$.
  
  The measure is defined as
  
  $$P_{BP} = \frac{1}{|C|}\sum_{s \in C}\frac{d_{I}(s)}{d_{C}(s)+d_{I}(s)} - 0.5,$$
  where $d_C$ is the number of edges between the node $s$ and nodes in $C$ and $d_I$ is the number of edges between the same node $s$ and nodes in $I$. The score is normalized by the cardinality of set of boundary nodes $C$. The values of $P_{BP}$ range from -0.5 to 0.5, where 0.5 indicates maximum polarization. Non-polarized network is expected to have values close to zero, whereas negative values indicate that the boundary nodes of community $A$ are more likely to connect to the other side $B$ than to own side.

  \medskip
  \item Dipole Polarization
  
  This measure applies label propagation for quantifying the distance between the influencers of each side. Its intuition is that a network is perfectly polarized when divided in two communities of the same size and opposite opinions. First, the top-k\% highest-degree nodes from both sides are selected. These nodes are assigned the "extreme opinion scores" of -1 or 1 depending on which side they belong to. For the influencer nodes, the $r_{t}$ is fixed to its the extreme value for all steps $t$. All the other nodes begin with a neutral opinion score $r_{t=0} = 0$. The opinion scores $r$ of the rest of nodes in the network are then updated by label propagation as follows
  
  $$ r_{t}(s) = \frac{\sum_{v} W_{sv}r_{t-1}(s)}{d(s)},$$
  
  where $W_{sv} = 1$ if there is an edge between the nodes $s$ and $v$, $r_{t-1}(s)$ is the opinion score of node $s$ at previous step and $d(s)$ is the degree of node $s$. This process is repeated until the convergence of opinion scores.
  
  Denote the average or the gravity center of positive and negative opinion scores by $gc^+$ and $gc^-$. The distance between the means of the opposite opinion score distributions is then $d = \frac{|gc^+ - gc^-|}{2}$. For the final polarization score $P_{DP}$, the distance $d$ is multiplied by $(1-\Delta a)$ to penalize the potential difference in the community sizes. The $\Delta a$ can be obtained either by a) taking the difference of definite integrals of the opposite opinion score distributions or b) by computing the absolute difference of the normalized community sizes. The latter is simply obtained by  $\Delta a = \frac{|n^+ - n^-|}{|A\cup B|}$, where $n^+$ denotes the number of nodes having a positive opinion score and $n^-$ denotes the number of nodes having a negative opinion score.
  
  The final polarization is calculated by
  
  $$P_{DP} = (1-\Delta a)d.$$
 
  The value of $P_{DP}$ can only be its maximum when the label-propagation based communities have equal sizes. The closer the means of opinion score distributions of both communities are, the lower polarization.

  \medskip
  \item Modularity
  
  Modularity is one of the most popular scores for describing discrepancy in a social network. Modularity measures how different the communities are from the corresponding communities of the ensemble of random graphs obtained by configuration model. The polarization based on modularity is simply the formula of modularity that is used to evaluate the quality of communities.
  
  $$P_Q = \frac{1}{2|E|}\sum_{ij}(W_{ij}-\frac{k_i k_j}{2|E|})\delta(c_i, c_j),$$
  
  where $|E|$ is the number of edges, $W_{ij}$ is the element of adjacency matrix and $k_i$ is the degree of node $i$. The value of $\delta(c_i, c_j)$ equals to one only when the nodes $i$ and $j$ belong to the same community, otherwise it is zero.
  
  \medskip
  \item E-I Index
  
  This simple measure, also known as \textit{Krackhardt E/I Ratio}, computes the relative density of internal connections within a community compared to the number of connections that community has externally. For two communities, it can be defined as
  
  $$P_{EI} = \frac{|C|}{|C'|},$$
  
  where $C$ is the cut-set $\{(s,t) \in E | s \in A, t \in B\}$ and $C'$ is the complement of that set $(C' = E/C)$.
  
  \medskip
  \item Adaptive E-I Index
  
  This measure is an extension of E-I Index as it accounts for different community sizes by using the density of links within each community. The Adaptive E-I Index becomes E-I Index when both of the communities have equal number of nodes. The measure is defined as
  
  $$P_{AEI} =  \frac{\sigma_{AA}+\sigma_{BB}-(\sigma_{AB}+\sigma_{BA})}{\sigma_{AA}+\sigma_{BB}+(\sigma_{AB}+\sigma_{BA})},$$
  
  where $\sigma_{AA}$ is the ratio of actual and potential links within the community $A$ (similary for $\sigma_{BB}$) and $\sigma_{AB}$ is the observed number of links between the communities $A$ and $B$ divided by the number of all potential links.
  
\end{enumerate}

\section{Additional Figures and Analysis}\label{C}
In this appendix, we include figures that summarize results of additional analysis. In section \ref{sec:alt}, we include alternative illustrations and additional analysis to support our arguments. In section \ref{sec:part}, we include results obtained using different partitioning methods.

\subsection{Alternative visualisations and additional analysis}\label{sec:alt}
\begin{figure}[H]
    \centering
    \includegraphics[width=1\textwidth]{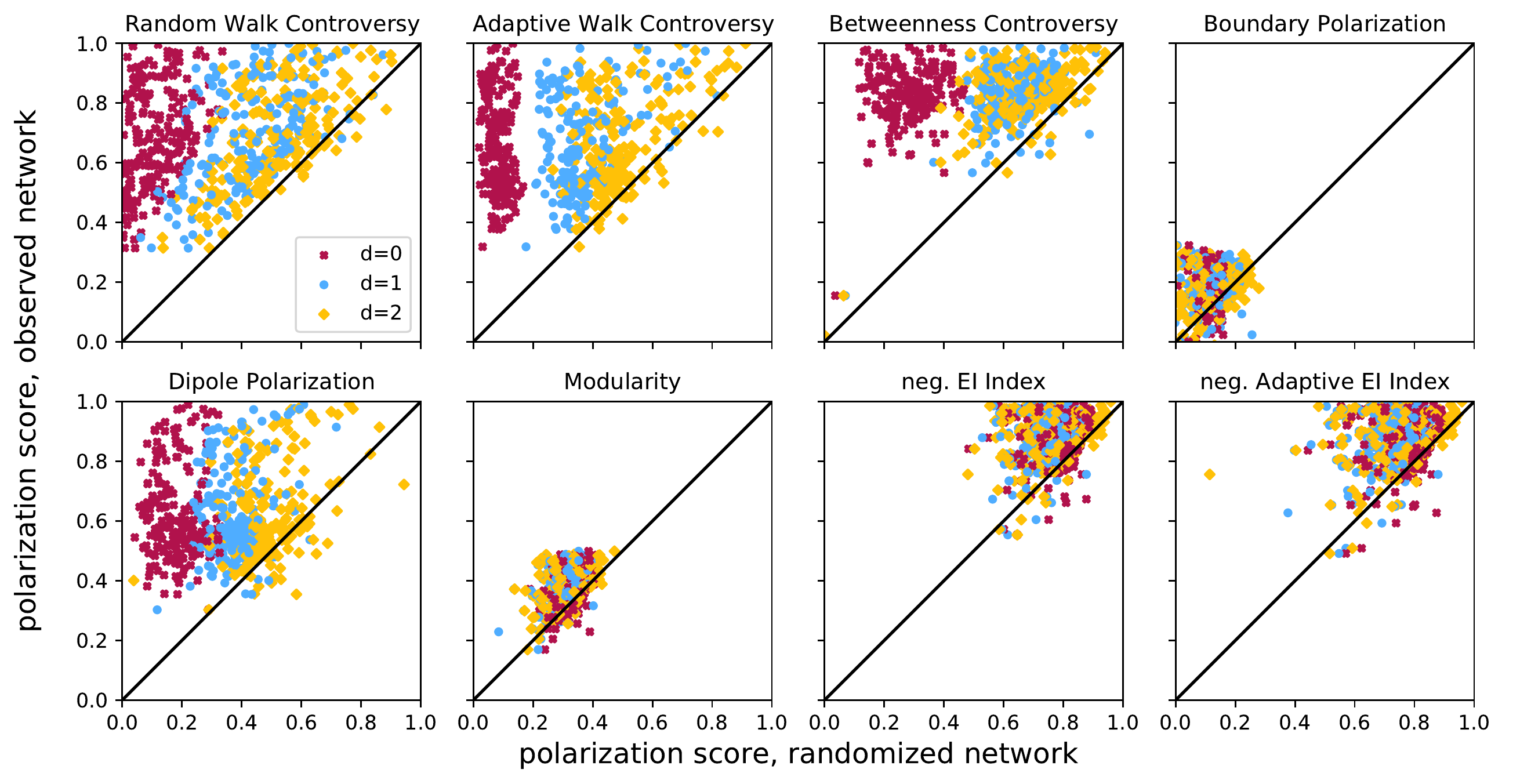}
    \caption{The observed polarization score as a function of the score computed for a randomized version of that network. Each point corresponds to the average polarization score of 500 randomizations of the original network. The colors represent the three different random models (see the legend in upper left panel).  
    Note that the negative values of Boundary Polarization have been left out in this visualisation. Figure contains data for all the 203 networks. Error bars for single points are ommitted because they are mostly too small to be visible.}
    \label{fig:dkmatrix150}
\end{figure}

\begin{figure}[H]
\centering
\includegraphics[width=1\textwidth]{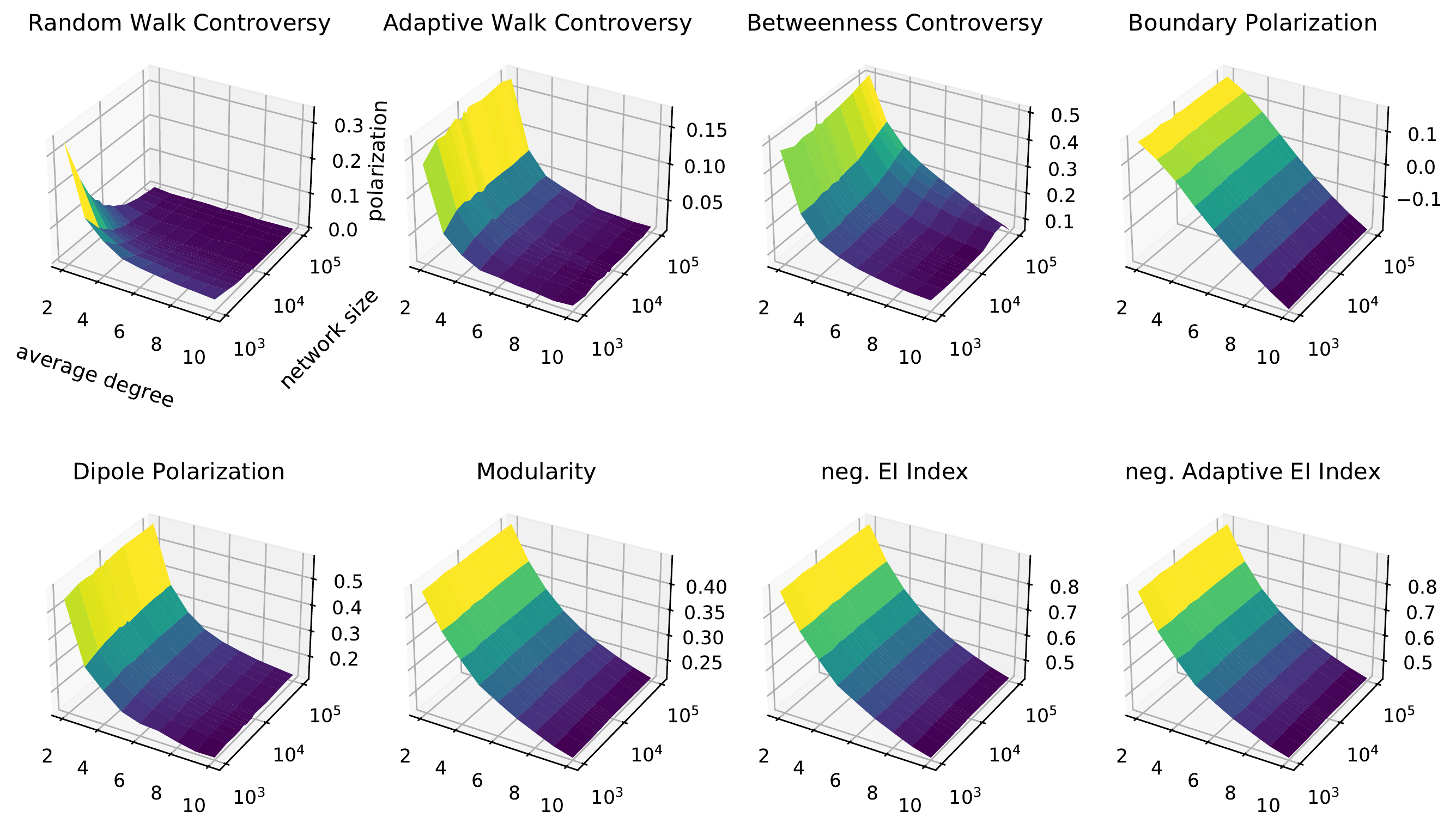}

\caption{3D plots displaying the effect of network size and average degree on polarization. For ER-networks, only RWC gives lower polarization as network grows but fixed parameter is required then. Many of the scores approaches zero polarization only after becoming very dense.}
\label{fig:3dplot}
\end{figure}

Fig. \ref{fig:dkmatrix150} is an alternative plot for illustrating our observation in Section \ref{rnetana}. Fig. \ref{fig:3dplot} is an alternative plot for illustrating our observation in Section \ref{densitysection}.

\begin{figure}[b]
    \centering
    \includegraphics[width=1\textwidth]{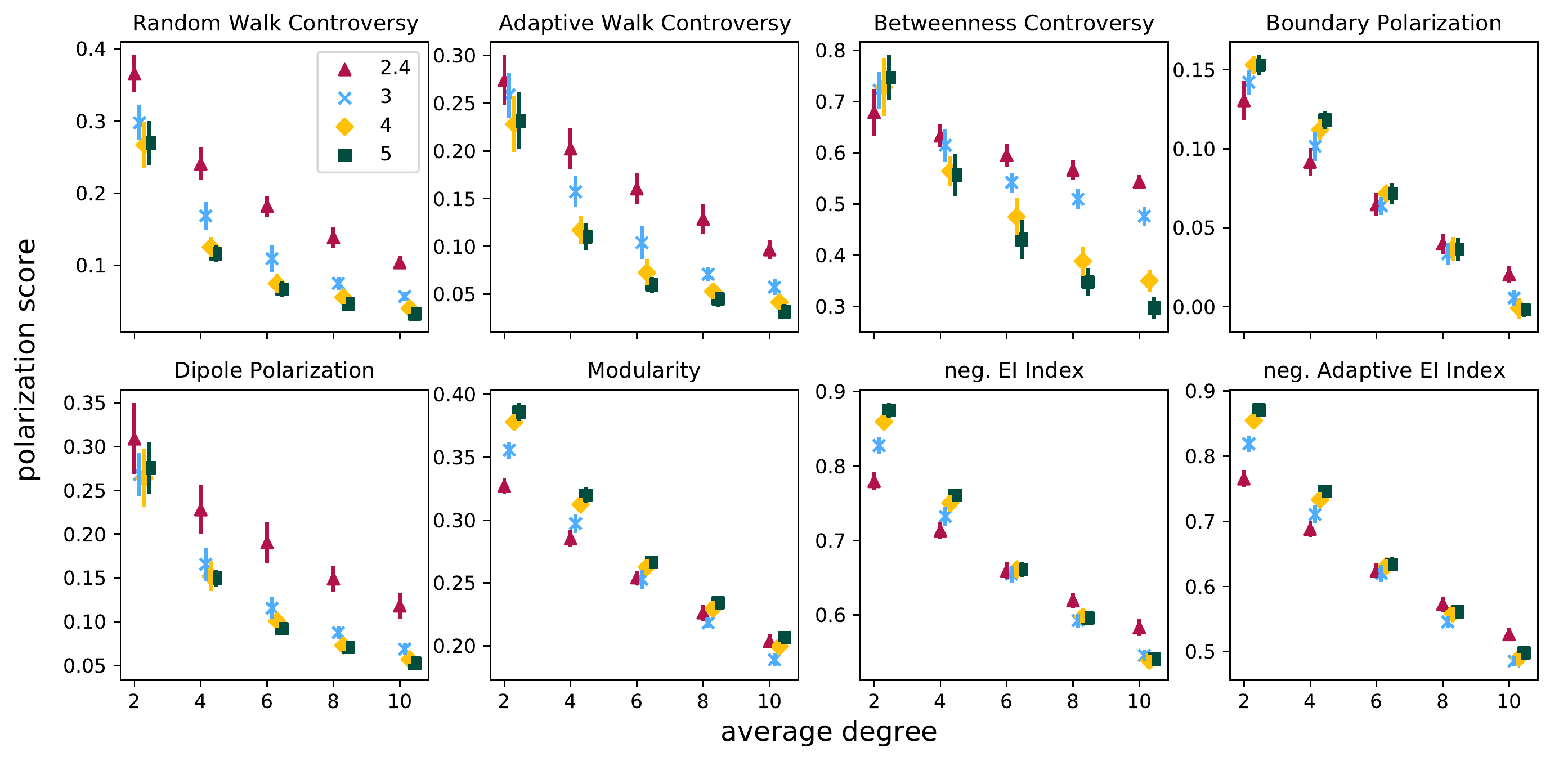}
    \caption{The effect of degree heterogeneity to polarization scores for the 8 scores in simulated scale-free networks with 2000 nodes. We show the expected polarization score and its standard deviation as a function of average degree. The symbols correspond to different exponent $\gamma$ values as indicated by the legend. The RWC values are lower for larger networks which was also shown to be the case for ER-networks in Section \ref{densitysection}. The rest of the scores are not dependent on network size. See Fig. \ref{fig:deghet1k} in the main text for the network with 1000 nodes}
    
    \label{fig:deghet2k}
\end{figure}

\begin{figure}
    \centering
    \includegraphics[width=1\textwidth]{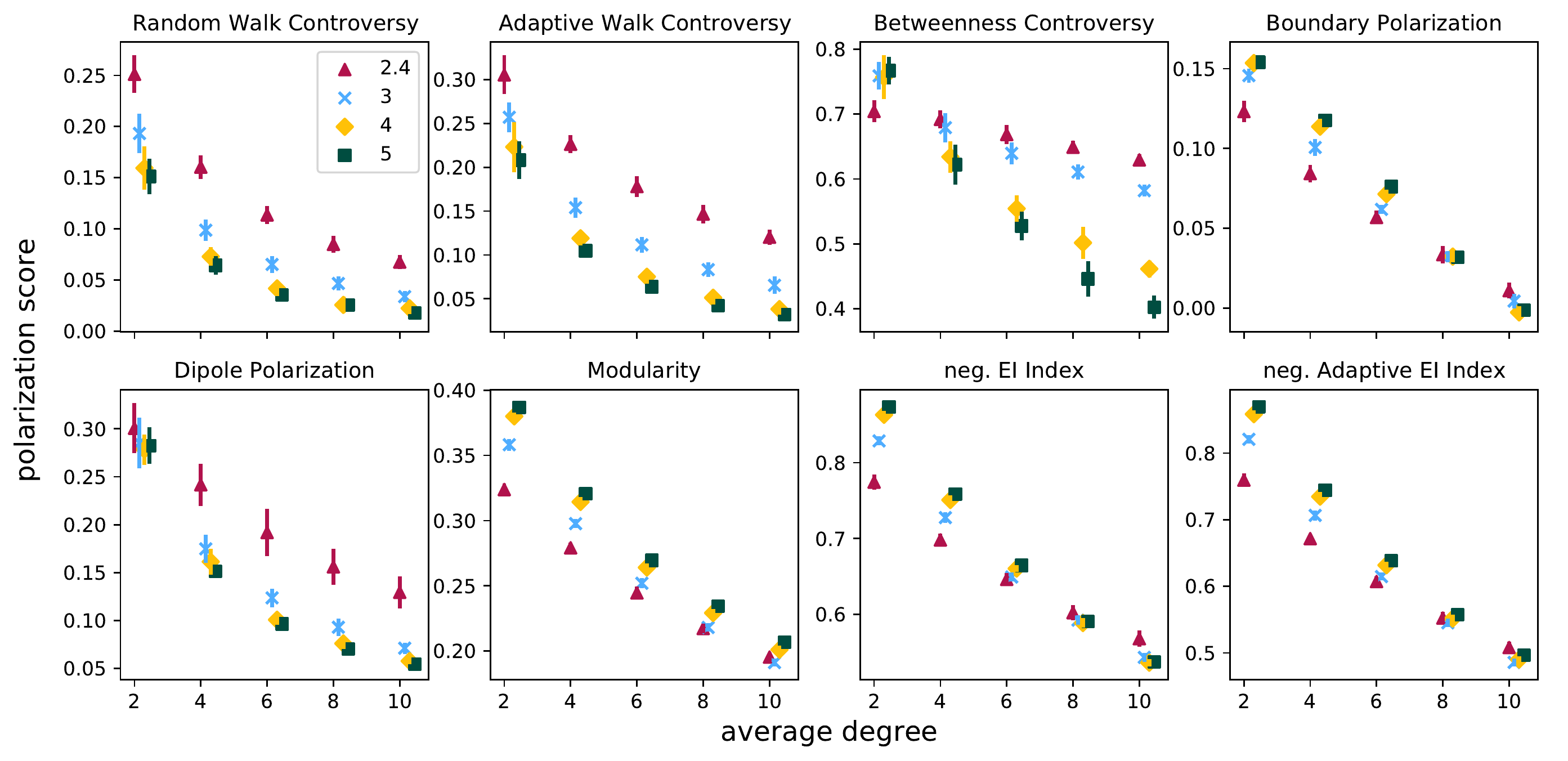}
    \caption{The effect of degree heterogeneity to polarization scores for the 8 scores in simulated scale-free networks with 5000 nodes. We show the expected polarization score and its standard deviation as a function of average degree. The symbols correspond to different exponent $\gamma$ values as indicated by the legend. See Fig. \ref{fig:deghet1k} in the main text for the network with 1000 nodes.}
    
    \label{fig:deghet5k}
\end{figure}

\begin{figure}
    \centering
    \includegraphics[scale=0.4]{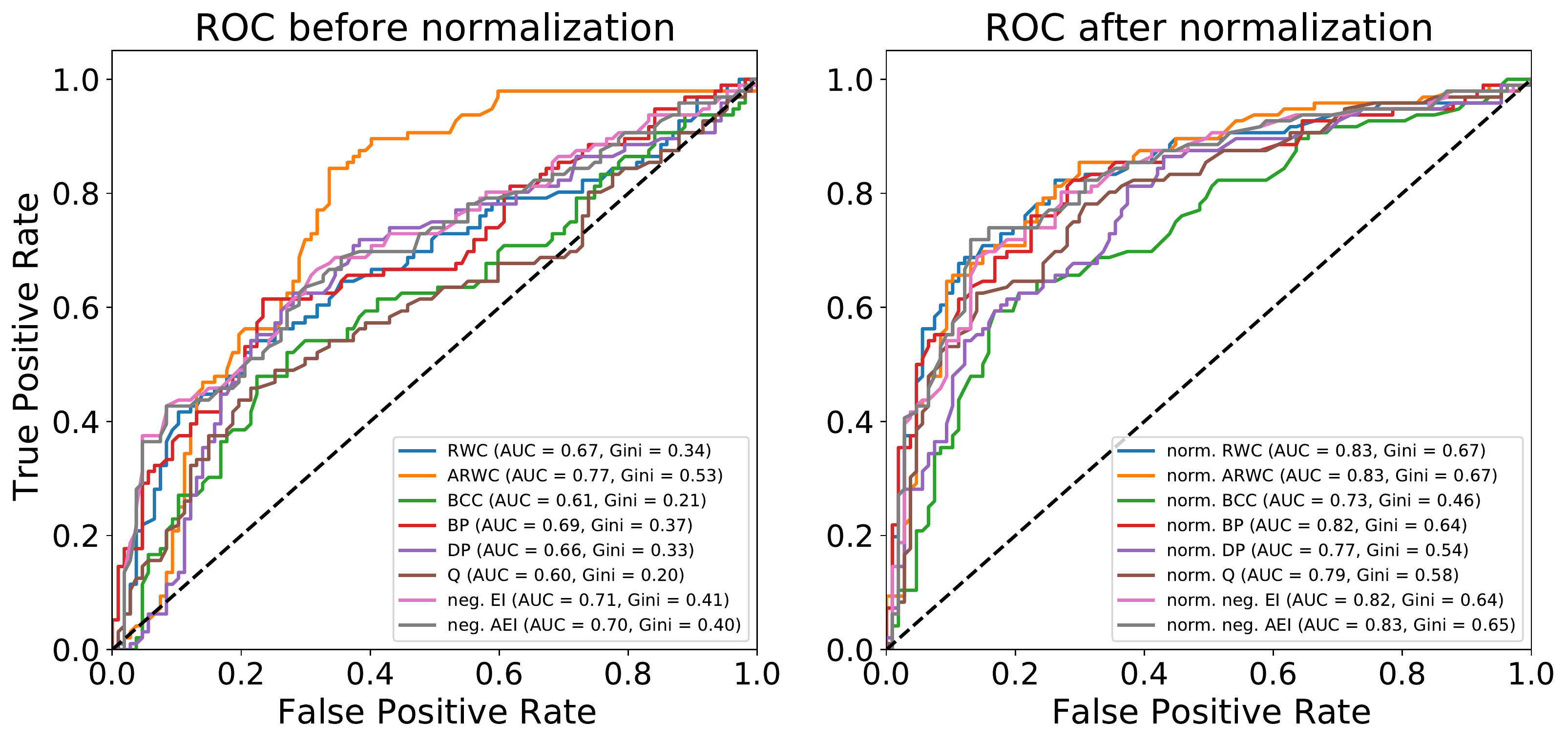}
    \caption{ROC curves, Gini coefficient values, and AUC values for the task of predicting manually curated labeling of polarized and non-polarized networks. The results shown (left) for the score values before the normalization and (right) after the normalization with standardized and denoised scores $\hat{\Phi}_{z}(G)$  (See Section \ref{normsection}).
Figures are based on all the 203 empirical networks described in section \ref{datasection}. The ROC curves for denoised polarization scores are in Fig. \ref{fig:roc}.}
    \label{fig:ROCXX}
\end{figure}

\begin{figure}
    \centering
    \includegraphics[width=1\textwidth]{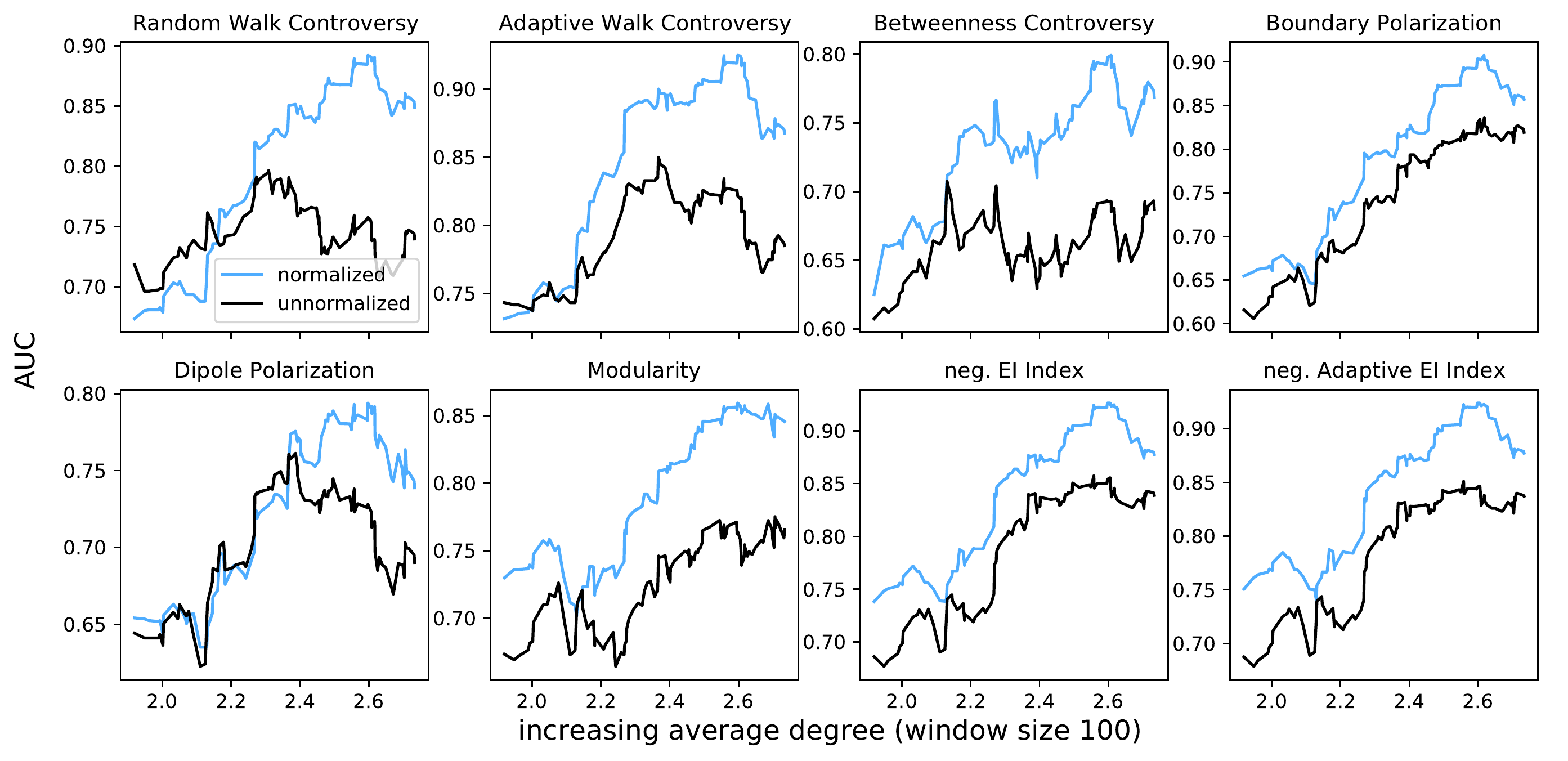}
    \caption{Quantifying how network's average degree affects the performance. We group the data such that there are 100 networks with consecutive degrees in our data, and create a set of such windows by varying the degree range.
    We then evaluate the AUC for the moving window of 100 networks. The plots show how the performances of both normalized and unnormalized scores become higher as the average degree increases for all the polarization methods. However, the normalization still improves the overall accuracy, especially for the less sparse networks. For instance, the normalization of RWC score improves the AUC approximately 0.10 units for networks with an average degree of 2.4 or higher.}
    \label{fig:new_aucdeg}
\end{figure}

\begin{figure}
    \centering
    \includegraphics[width=1\textwidth]{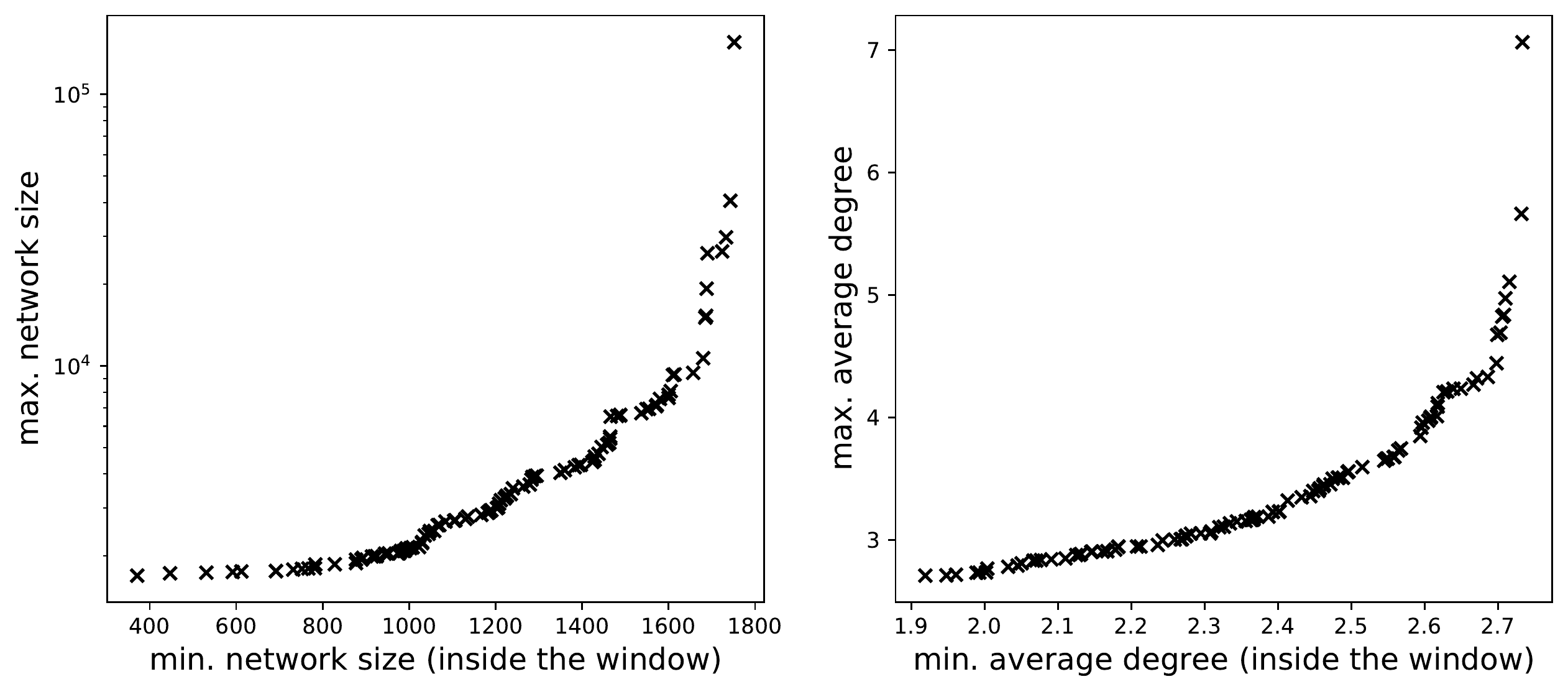}
    \caption{Additional information about the windows for Fig. \ref{fig:new_aucsize} and Fig. \ref{fig:new_aucdeg}. The smallest value of the window is on x-axis and, respectively, the largest value of the window is located on y-axis.}
    \label{fig:new_aucdeg_scale}. 
\end{figure}

\begin{figure}
    \centering
    \includegraphics[width=1\textwidth]{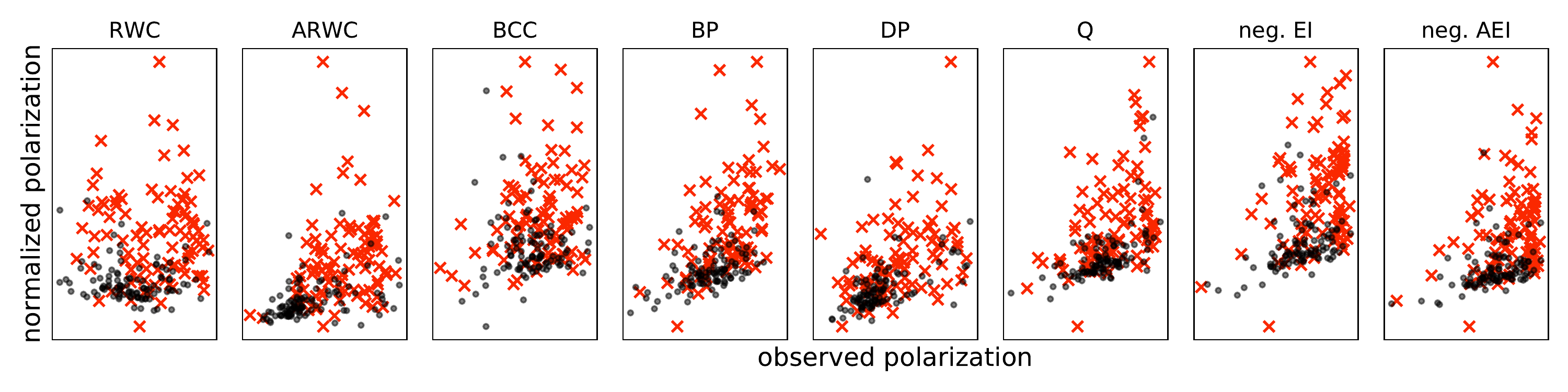}
    \caption{Similar scatter plot as in Fig. \ref{fig:normpanel} but for denoised polarization scores with standardization. Red crosses are networks labeled as polarized and black points are networks labeled as non-polarized. An outlier was removed from the plots. 
    Note that the scales for the scores are different and not shown in this figure.}
    \label{fig:scatterstandard}
\end{figure}

The effects of heterogeneous degree sequences on polarization measures were also studied for larger networks (Figs. \ref{fig:deghet2k} and \ref{fig:deghet5k}). To see the performance stability across networks with different average degrees, see Figs. \ref{fig:new_aucdeg} and \ref{fig:new_aucdeg_scale}. Finally, the Figs. \ref{fig:ROCXX} and \ref{fig:scatterstandard} show the performance of the standardized denoised polarization scores.

\subsection{Results for alternative graph partitioning methods}\label{sec:part}
\begin{figure}[htbp]
    \centering
    \includegraphics[width=1\textwidth, trim = {0.275cm 0 0.255cm 0}, clip]{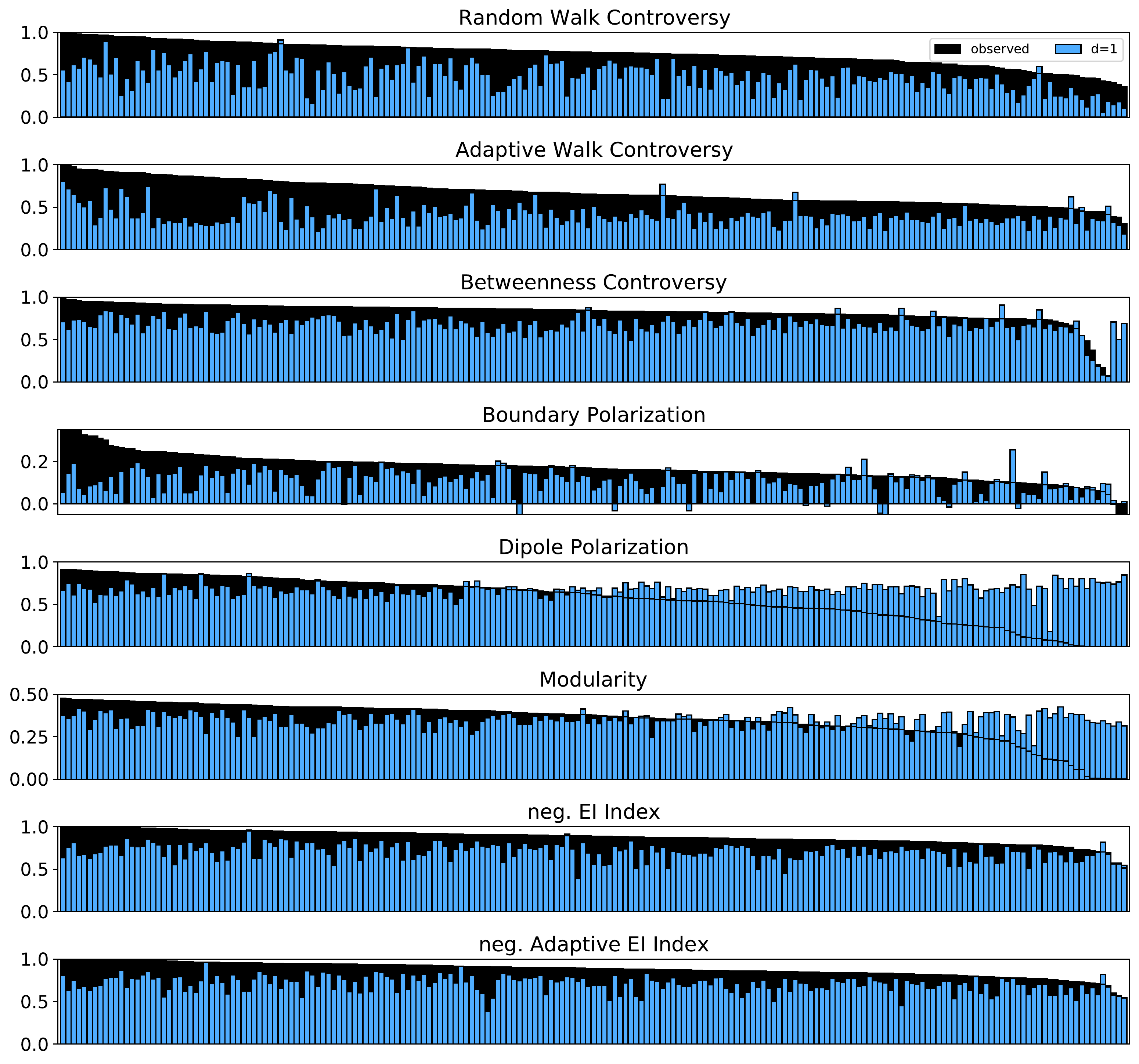}
    \caption{Polarization scores for the 203 observed networks and their shuffled versions. Each bar corresponds a score, and scores for a network and its randomized versions are on top of each other: observed network  is represented with black bars and scores competed for random networks where degree sequence is preserved ($d=1$) are shown in blue. An interpretation for the figure is that, the amount of blue that is shown tells how much of the total bar height (the score value) is explained by the degree distribution and the amount of black that is shown is not explained by it. Note that in some cases, the randomized networks produce higher scores than the original network and in this case the black bar is fully covered by the blue bar. In this case we draw a black horizontal line on top of the blue bar indicating the height of the black bar. The difference of this figure to Fig.~\ref{fig:bar} in the main text is that groups in this figure are produced with spectral clustering and only the null model for degree sequence is shown.
    }
    \label{fig:bar_spectral}
\end{figure}

\begin{figure}[htbp]
    \centering
    \includegraphics[width=1\textwidth, trim = {0.275cm 0 0.255cm 0}, clip]{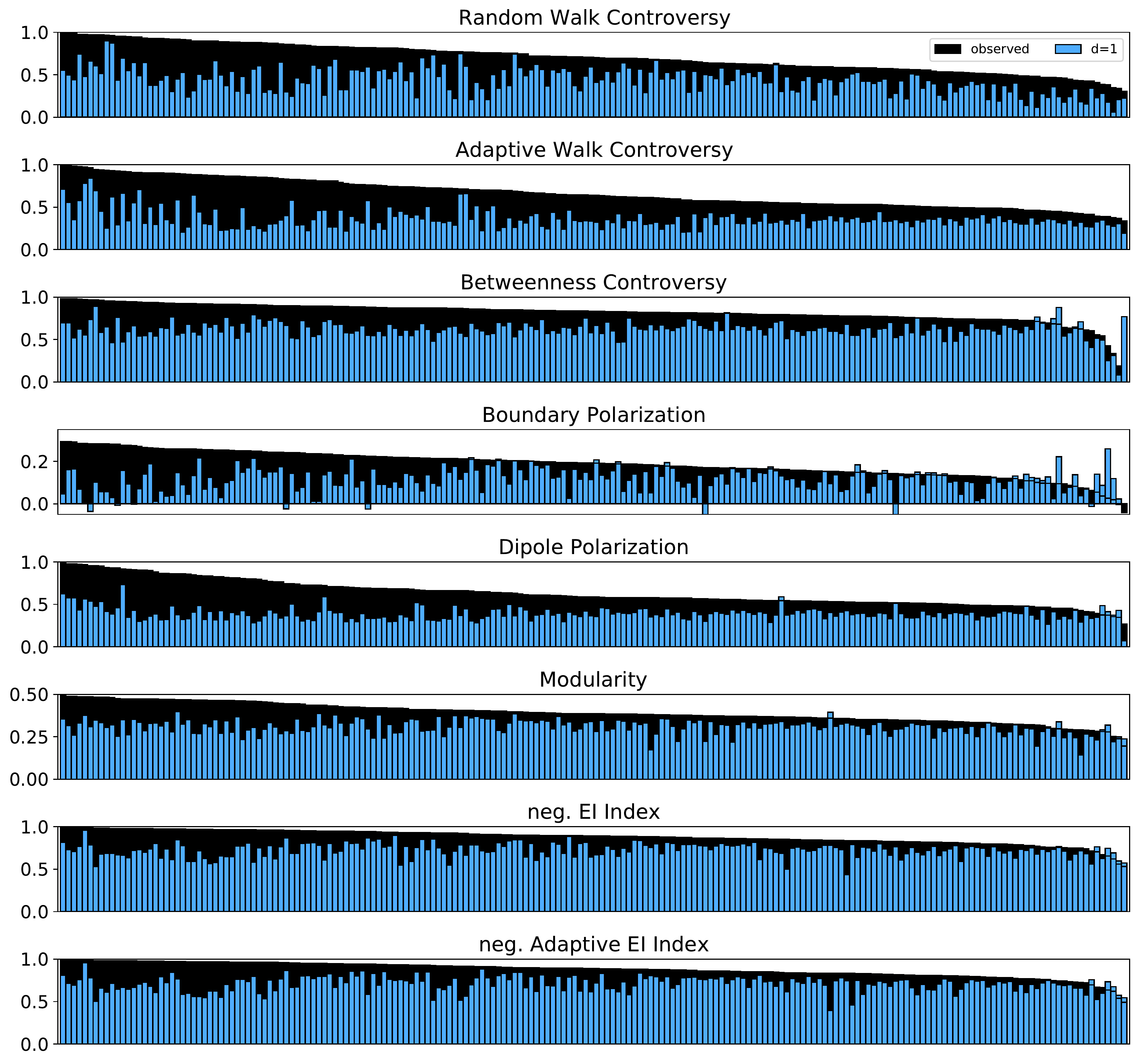}
    \caption{Polarization scores for the 203 observed networks and their shuffled versions. Each bar corresponds a score, and scores for a network and its randomized versions are on top of each other: observed network  is represented with black bars and scores competed for random networks where degree sequence is preserved ($d=1$) are shown in blue. An interpretation for the figure is that, the amount of blue that is shown tells how much of the total bar height (the score value) is explained by the degree distribution and the amount of black that is shown is not explained by it. Note that in some cases, the randomized networks produce higher scores than the original network and in this case the black bar is fully covered by the blue bar. In this case we draw a black horizontal line on top of the blue bar indicating the height of the black bar. The difference of this figure to Fig.~\ref{fig:bar} in the main text is that groups in this figure are fine tuned with modularity optimization and only the null model for degree sequence is shown.
    }
    \label{fig:bar_maxmod}
\end{figure}

\begin{figure}
    \centering
    \includegraphics[width=1\textwidth]{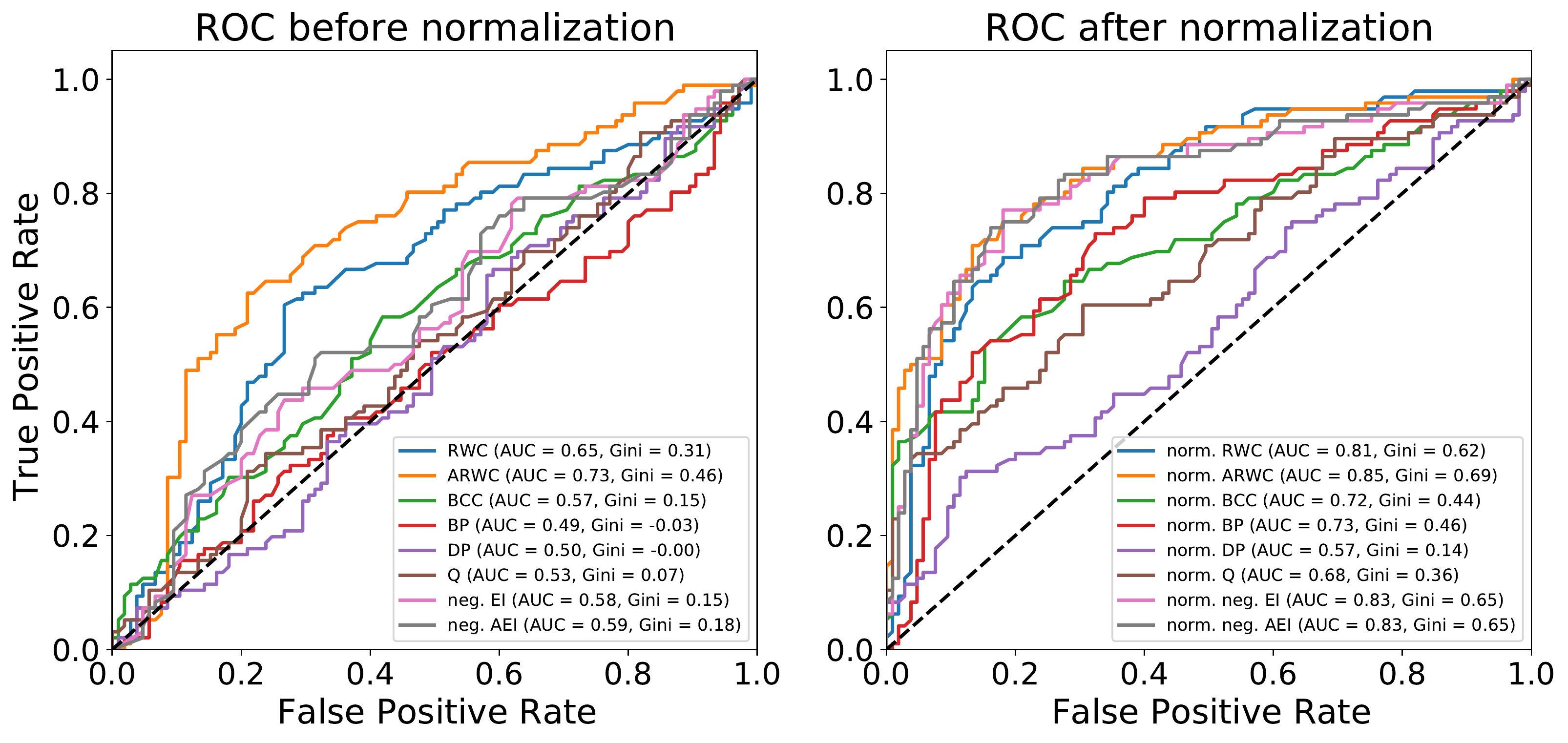}
    \caption{
    ROC curves, Gini coefficient values, and AUC values for the task of predicting manually curated labeling of polarized and non-polarized networks. The results shown (left) for the score values before the normalization and (right) after the normalization with denoised scores $\hat{\Phi}(G)$  (see main text). Figures are based on all the 203 empirical networks described in section \ref{datasection}. 
The difference of this figure to Fig.~\ref{fig:roc} in the main text is that groups in this figure are produced with spectral clustering.}
    \label{fig:ROC_SPECTRAL}
\end{figure}

\begin{figure}
    \centering
    \includegraphics[width=1\textwidth]{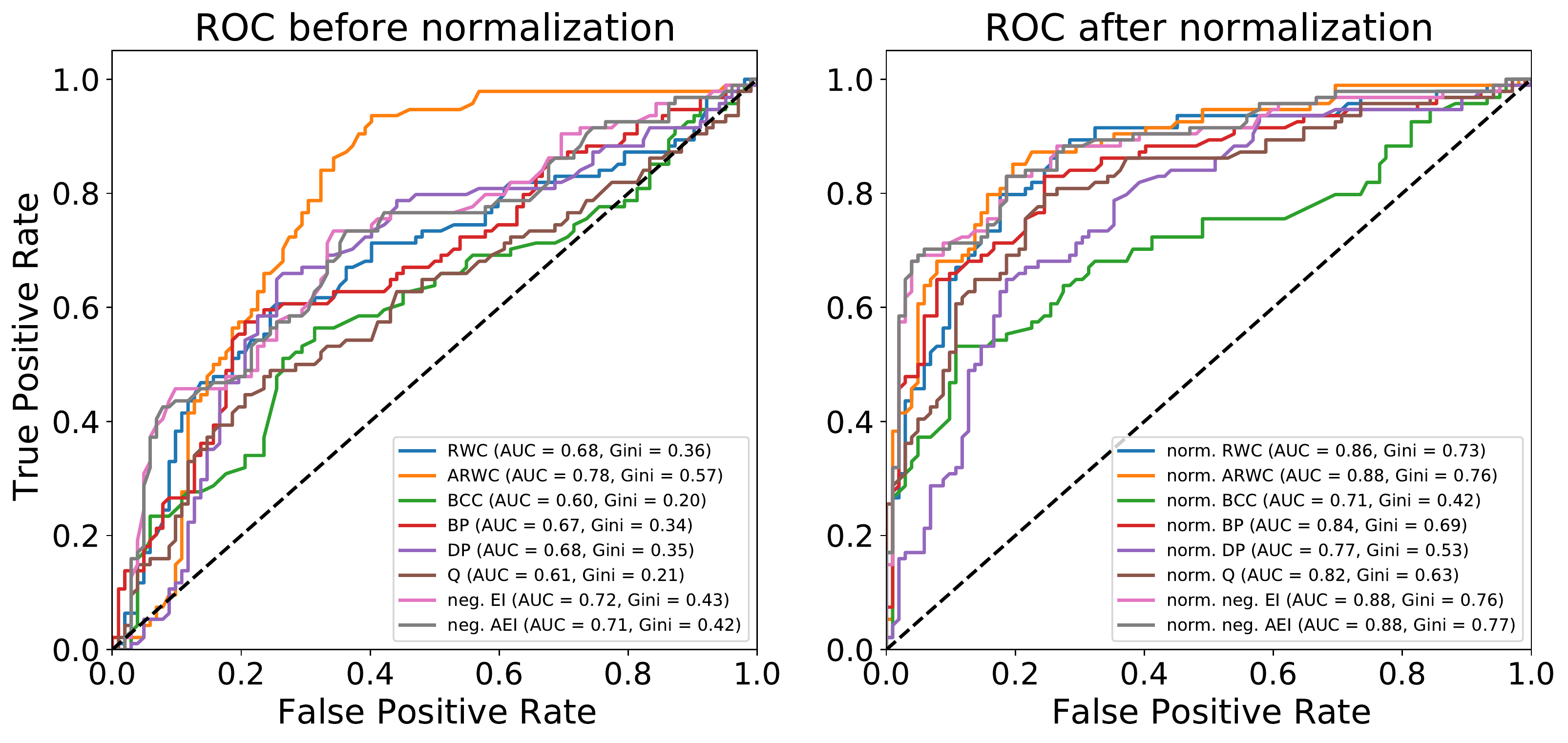}
    \caption{ROC curves, Gini coefficient values, and AUC values for the task of predicting manually curated labeling of polarized and non-polarized networks. The results shown (left) for the score values before the normalization and (right) after the normalization with denoised scores $\hat{\Phi}(G)$  (see main text). Figures are based on all the 203 empirical networks described in section \ref{datasection}. 
The difference of this figure to Fig.~\ref{fig:roc} in the main text is that groups in this figure are fine tuned with modularity optimisation.}
    \label{fig:ROC_MOD}
\end{figure}

In addition to METIS, we perfomed our analysis using two alternative clustering methods: regularized spectral clustering and modularity optimization. While the number of links between the two groups is used by METIS as an optimisation criteria, the intuition behind spectral clustering is related to finding groups where a random walker would remain in the starting cluster as long as possible. Further, modularity measures the excess fraction of links inside the groups as compared to a null model. The clusters obtained by METIS had already high modularity values. Therefore, for optimizing the modularity, we used the partition produced by METIS as the pre-partition for which a fine-tuning was performed: We then optimised the partition for maximum modularity by a greedy stochastic optimization method which consecutively tries to swap the cluster of a random node and accepts it if the value of the target function improves \cite{klin}. A reasonable convergence was achieved when the number of swaps was equal to two times the number of nodes in the network. Figures \ref{fig:bar_spectral} and \ref{fig:bar_maxmod} display the noise bar analysis equivalent to Fig. \ref{fig:bar} in the main text for these two additional methods, but only considering the configuration model ($d=1$).
Figures \ref{fig:ROC_SPECTRAL} and \ref{fig:ROC_MOD} are alternatives for Fig.~\ref{fig:roc} displaying the ROC curves related to the classification task presented in Sec.~\ref{sec:evaluation}. 
\clearpage

\end{document}